\documentclass[aps,twocolumn,prl,showpacs,floatfix,altaffilletter]{revtex4-1}
\usepackage{graphicx}
\usepackage{bm}
\usepackage{amsmath}
\usepackage{amsfonts}
\usepackage{amssymb}
\usepackage{epstopdf}
\usepackage{natbib} %\bibliographystyle{prsty}
\usepackage{xspace}
\newcommand{\fesc}{FeSc$_2$S$_4$\xspace}
\newcommand{\fep}{Fe$^{2+}$\xspace}
\begin{document}
\title{Magnetic field dependence of excitations near spin-orbital quantum criticality}
\author{A. Biffin,$^{1,2}$ Ch. R\"{u}egg,$^{1,3}$ J. Embs,$^1$ T. Guidi,$^4$ D. Cheptiakov,$^1$ A. Loidl,$^5$ V.
Tsurkan,$^{5,6}$ and R. Coldea$^2$}
\affiliation{$^{1}$Laboratory
for Neutron Scattering and Imaging, Paul Scherrer Institut, 5232
Villigen, Switzerland}
\affiliation{$^{2}$Clarendon Laboratory,
University of Oxford, Parks Road, Oxford, OX1 3PU, United Kingdom}
\affiliation{$^{3}$Department of Quantum Matter Physics,
University of Geneva, 1211 Geneva, Switzerland}
\affiliation{$^{4}$ISIS Facility, Rutherford Appleton Laboratory,
Chilton, Didcot, OX11 0QX, United Kingdom}
\affiliation{$^{5}$Experimental Physics 5, Center for Electronic
Correlations and Magnetism, Institute of Physics, University of
Augsburg, D-86159 Augsburg, Germany}
\affiliation{$^{6}$Institute
of Applied Physics, Academy of Sciences of Moldova, MD-2028,
Chisinau, Republic of Moldova} \pacs{}

\begin{abstract}
The spinel FeSc$_2$S$_4$ has been proposed to realize a
near-critical spin-orbital singlet (SOS) state, where entangled
spin and orbital moments fluctuate in a global singlet state on
the verge of spin and orbital order. Here we report powder
inelastic neutron scattering measurements that observe the full
bandwidth of magnetic excitations and we find that spin-orbital
triplon excitations of an SOS state can capture well key aspects
of the spectrum in both zero and applied magnetic fields up to
8.5~T. The observed shift of low-energy spectral weight to higher
energies upon increasing applied field is naturally explained by
the entangled spin-orbital character of the magnetic states, a
behavior that is in strong contrast to spin-only singlet ground
state systems, where the spin gap decreases upon increasing
applied field.
\end{abstract}
\maketitle

\begin{figure}[htbp]
\includegraphics[width=0.49\textwidth]{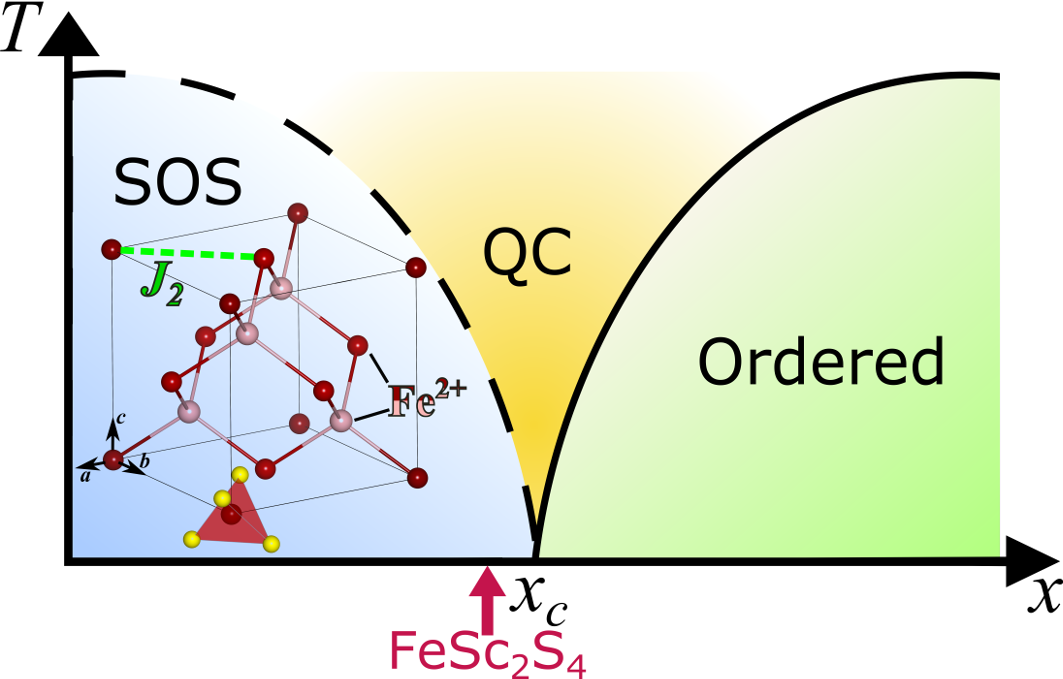}
\caption[]{(Color online). Generic phase diagram as a function of
$x=J_2/\lambda$ \cite{chen_prl} with location of \fesc indicated
by the vertical arrow. Labels SOS, QC and Ordered stand for
spin-orbital singlet, quantum critical, and spin and orbitally
ordered, respectively. Solid/dashed line indicates a phase
transition boundary/crossover. Inset shows \fep ions in \fesc are locally
coordinated by a regular tetrahedron (red shaded polyhedron) of
S$^{2-}$ atoms (yellow balls) and are arranged in two FCC
sublattices (light/dark red balls) with intra-sublattice NN AFM
exchange $J_2$. } \label{fig:structure}
\end{figure}

When magnetic ions posses an orbital degeneracy in addition to
spin, the combined effects of the on-site spin-orbit coupling and
the inter-site magnetic exchange interactions have been
theoretically proposed to stabilize correlated states with
entangled spin-orbital character and novel quasiparticles
\cite{chen_prl,excitonic_magnetism}. Generally such physics is not
directly experimentally accessible as symmetry-lowering
Jahn-Teller (JT) structural distortions \cite{JT} tend to lift
orbital degeneracy leaving a spin-only degree of freedom. However,
in the case of relatively strong spin-orbit coupling, or certain
crystal structures where JT distortions are inhibited by the
lattice geometry, spin-orbit entanglement can become
manifest. For $d^4$ \cite{excitonic_magnetism} and $d^6$
\cite{low} transition metal ions in certain high-symmetry crystal
environments the single-ion ground state is a spin-orbit entangled
$J_{\rm eff}=0$ singlet with an excited $J_{\rm
eff}=1$ triplet at higher energy. In this case, a theoretically-proposed phase
diagram \cite{chen_prl} as a function of the ratio $x$ of magnetic
exchange couplings to the singlet-triplet gap $\lambda$ is shown
in Fig.~\ref{fig:structure}. Cooperative spin and orbital
order is expected for $x>x_c$, with a novel amplitude (``Higgs'')
mode for $x \gtrsim x_c$ \cite{excitonic_magnetism,ruthenate} and
entangled spin-orbital fluctuations present at the critical point
$x_c$. In the regime of moderate exchange interactions,
$x \lesssim x_c$, spins and orbitals are expected to be strongly
fluctuating in a quantum paramagnetic state denoted as a
``spin-orbital singlet'' (SOS), with strong correlations between
sites \cite{chen_prl}. Even though the SOS state has no spin or
orbital order, it supports quasiparticles, so called
``spin-orbital triplons'' (or ``spin-orbitons''
\cite{mittelstadt}), corresponding to isotropically-polarized,
spin and orbital density wave packets that can propagate
coherently across the lattice.

%%%%%%%%%%%%%%%%%%%%%%%%%%%%%%%%%%%%%%%%%%%%%%%%%%%%%%%%%%%%%
%%%%%%%%%%%%%%%%%%%%%%%%%%%%%%%%%%%%%%%%%%%%%%%%%%%%%%%%%%%%%

The spinel \fesc has been proposed \cite{fritsch,krimmel,chen_prl}
as a unique candidate to display a SOS state with
intermediate-strength exchange interactions ($x \lesssim x_c$)
that bring it almost on the verge of spin and orbital order.
It is the only known system to explore the physics of
highly-dispersive spin-orbital triplons, that may be close to
spin-orbital quantum criticality. Here we report inelastic neutron
scattering (INS) measurements over the full bandwidth of the
magnetic excitations and we find good agreement with the expected
spectrum of spin-orbital triplons of a near-critical SOS state. In
applied magnetic field we observe a striking shift of the
low-energy spectral weight to higher energies, a direct fingerprint of the entangled
spin-orbital character of the magnetic states.

\fesc has a cubic crystal structure with space group $Fd\bar{3}m$
(no. 227) and lattice parameter $a=10.51$~\AA{} at 300~K
\cite{tomas}. \fep ions are tetrahedrally-coordinated by S$^{2-}$
and in this crystal field of cubic symmetry the one-electron $d$
orbital states of \fep are split into a lower $e$-doublet and
upper $t_2$-triplet. Hund's coupling stabilizes a high-spin
($S=2$) state, $e^3t_2^3$, with a two-fold orbital degeneracy. The
atomic spin-orbit interaction $\lambda_0 \bm{L} \cdot \bm{S}$
lifts this two-fold orbital and five-fold spin degeneracy to
stabilize a SOS ground state with wavefunction \cite{low}
\begin{equation} \label{eq:gsideal}
\frac{1}{\sqrt{2}}|3z^2-r^2\rangle|0\rangle+
\frac{1}{2}|x^2-y^2\rangle\left(|-2\rangle+|+2\rangle\right),
\end{equation}
where for each term the first ket gives the (multi-electron)
orbital state and the second ket the $S_z$ eigenvalue. The first
excited state is a triplet above a gap $\lambda$ and local
singlet-triplet transitions then form the key
ingredient from which coherently-propagating triplons develop in
the presence of inter-site exchange interactions.

Previous susceptibility, specific heat and NMR measurements on
\fesc \cite{buttgen, fritsch} showed no clear anomalies indicative
of spin or orbital order in spite of strong magnetic interactions
manifested by a large antiferromagnetic (AFM) Curie-Weiss
temperature of $-45$~K, indicating that the material may indeed be in the SOS phase. INS
studies \cite{krimmel} focusing on the very low energy dynamics
indicated that the dominant magnetic interaction is an AFM exchange
$J_2$ between spins located at next-nearest neighbor (NNN) sites.
This splits the diamond lattice into two magnetically-decoupled,
frustrated FCC lattices (light/dark sites in
Fig.~\ref{fig:structure}), where $J_2$ acts on NN bonds.

%%%%%%%%%%%%%%%%%%%%%%%%%%%%%%%%%%%%%%%%%%%%%%%%%%%%%%%%%%%%%
\begin{figure}[!htbp]
\includegraphics[width=0.4\textwidth]{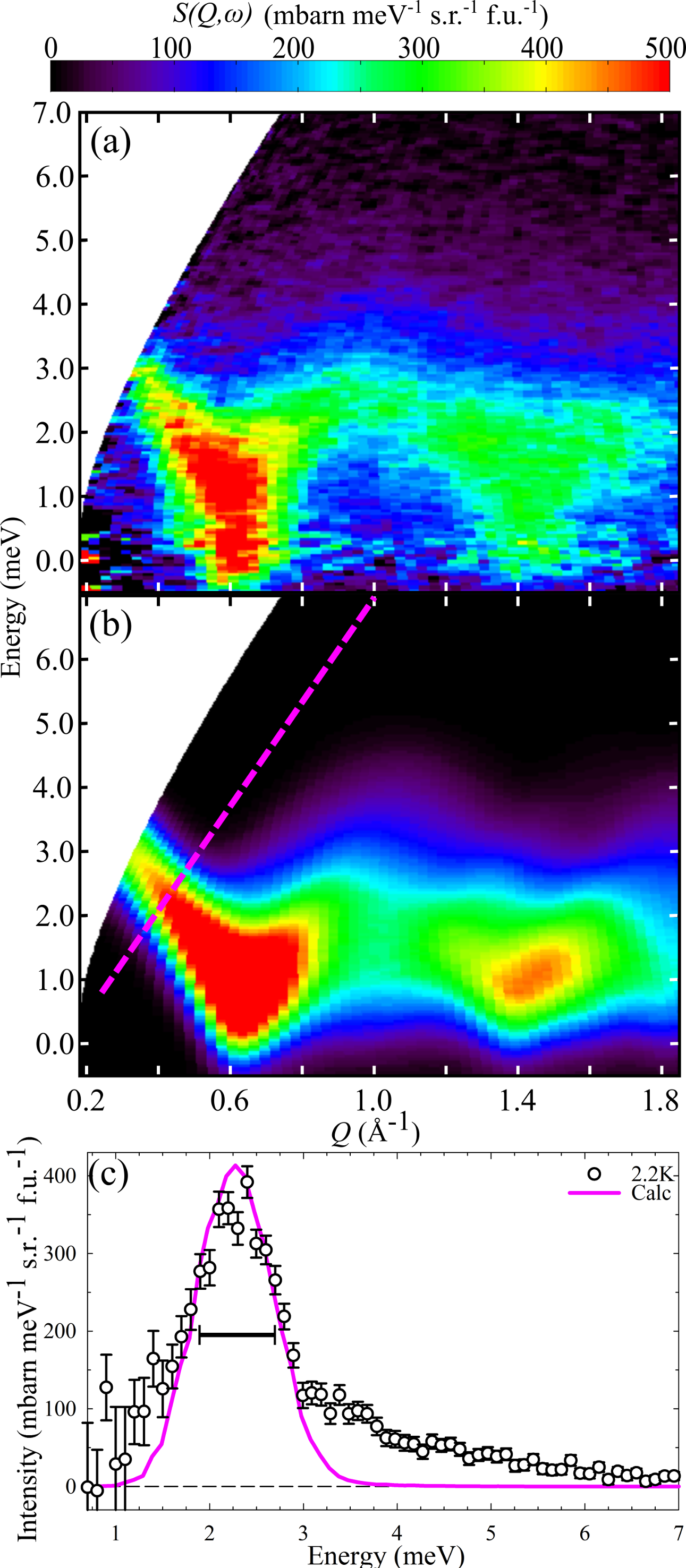}
\caption[]{(Color online). (a) Background subtracted (see \cite{supplemental}) zero field INS data observing highly
dispersive magnetic excitations (MERLIN, $E_i$=15~meV, 12~h
counting time, energy resolution 0.79~meV FWHM on the elastic
line). (b) Corresponding
one-triplon spectrum including convolution with the 
instrumental resolution. (c) Intensity along the tilted dashed
line direction in (b), compared to the model calculation (solid
line). Horizontal bar indicates expected peak FWHM due to
instrumental resolution and spherical averaging.}\label{fig:merlin}
\end{figure}
%%%%%%%%%%%%%%%%%%%%%%%%%%%%%%%%%%%%%%%%%%%%%%%%%%%%%%%%%%%%%

\begin{figure}[!htbp]
\begin{center}
\includegraphics[width=0.4\textwidth]{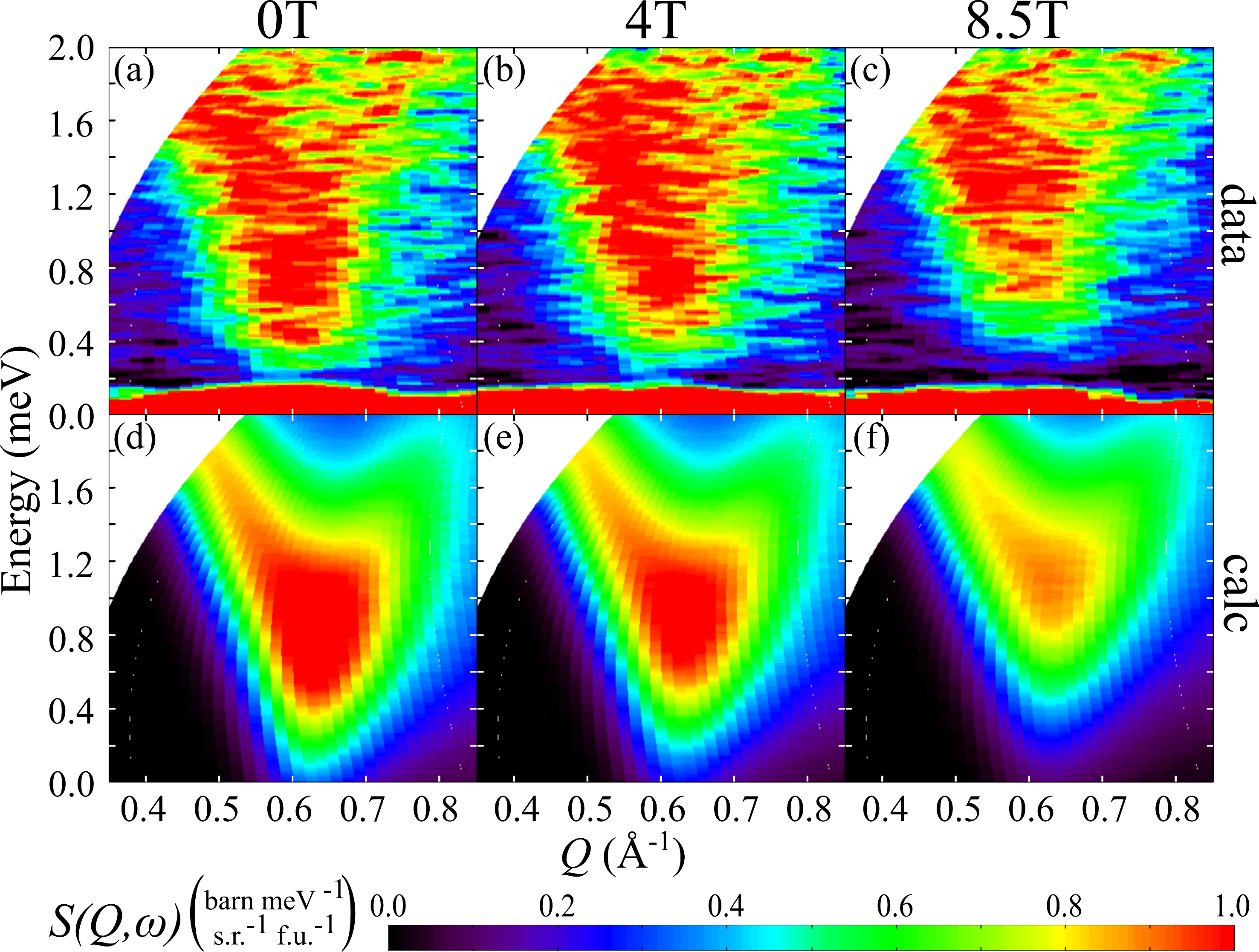}
\caption[]{(Color online) (a-c) Background-subtracted INS data at 0, 4 and 8.5~T
magnetic field compared with model calculations (d-f),
respectively. The data was collected using FOCUS with
$E_i=3.27$~meV, elastic line energy resolution 0.18~meV FWHM and
11~h counting per setting.} \label{fig:focus}
\end{center}
\end{figure}

We have probed the magnetic excitations using INS measurements
first in zero magnetic field and at temperatures 2.2-50~K using
the direct-geometry, time-of-flight spectrometer MERLIN at the
ISIS neutron source \cite{merlin}. The sample was a 4~g powder of
\fesc synthesized as described in \cite{supplemental} and used in
previous thermodynamic and diffraction studies \cite{fritsch}. The
INS intensities were converted into absolute units by
normalization to data measured on a vanadium standard. For
incident neutrons of energy $E_i=15$~meV the covered phase space
observed the full bandwidth of magnetic excitations, which showed
prominent dispersions with a bandwidth extending to around 4~meV
at the lowest temperatures, as shown in Fig.~\ref{fig:merlin}(a).
The high-temperature data was used to parameterize and subtract
the non-magnetic background (as described in \cite{supplemental}),
such that Fig.~\ref{fig:merlin}(a) shows the magnetic signal only.
Within experimental uncertainty no additional magnetic transitions
were detected at higher energy transfers (data collected using
incident neutron energies up to 200~meV). This is consistent with the
expectation that the single-ion ground state is close to the SOS
wavefunction in (\ref{eq:gsideal}), for which no other
(crystal-field) transitions are symmetry allowed
\cite{supplemental}. In agreement with previous low-energy studies
\cite{krimmel}, we observe a softening of the magnetic excitations
near a critical wavevector $Q_S \approx 0.6$~\AA$^{-1}$ [see Fig.~\ref{fig:merlin}(a)],
whose magnitude coincides with the structurally-forbidden $(100)$
reciprocal lattice position (in units of $2\pi/a$) and a natural
wavevector for AFM ordering on the FCC lattice \cite{chen_prl}.
Higher-resolution measurements shown in Fig.~\ref{fig:focus}(a)
indicate a clear suppression of scattering weight below
$\sim$0.4~meV, indicating that the gap is much smaller
than the full bandwidth of the magnetic excitations extending to
around 4~meV. This is consistent with the proposal that \fesc is
in the very close proximity of the critical point between SOS and
magnetic/orbital order, at which the gap would be expected to close
\cite{chen_prl}.

%%%%%%%%%%%%%%%%%%%%%%%%%%%%%%%%%%%%%%%%%%%%%%%%%%%%%%%%%%%%%
%%%%%%%%%%%%%%%%%%%%%%%%%%%%%%%%%%%%%%%%%%%%%%%%%%%%%%%%%%%%%

%%%%%%%%%%%%%%%%%%%%%%%%%%%%%%%%%%%%%%%%%%%%%%%%%%%%%%%%%%%%%
\begin{figure}[!htbp]
\begin{center}
\includegraphics[width=0.4\textwidth]{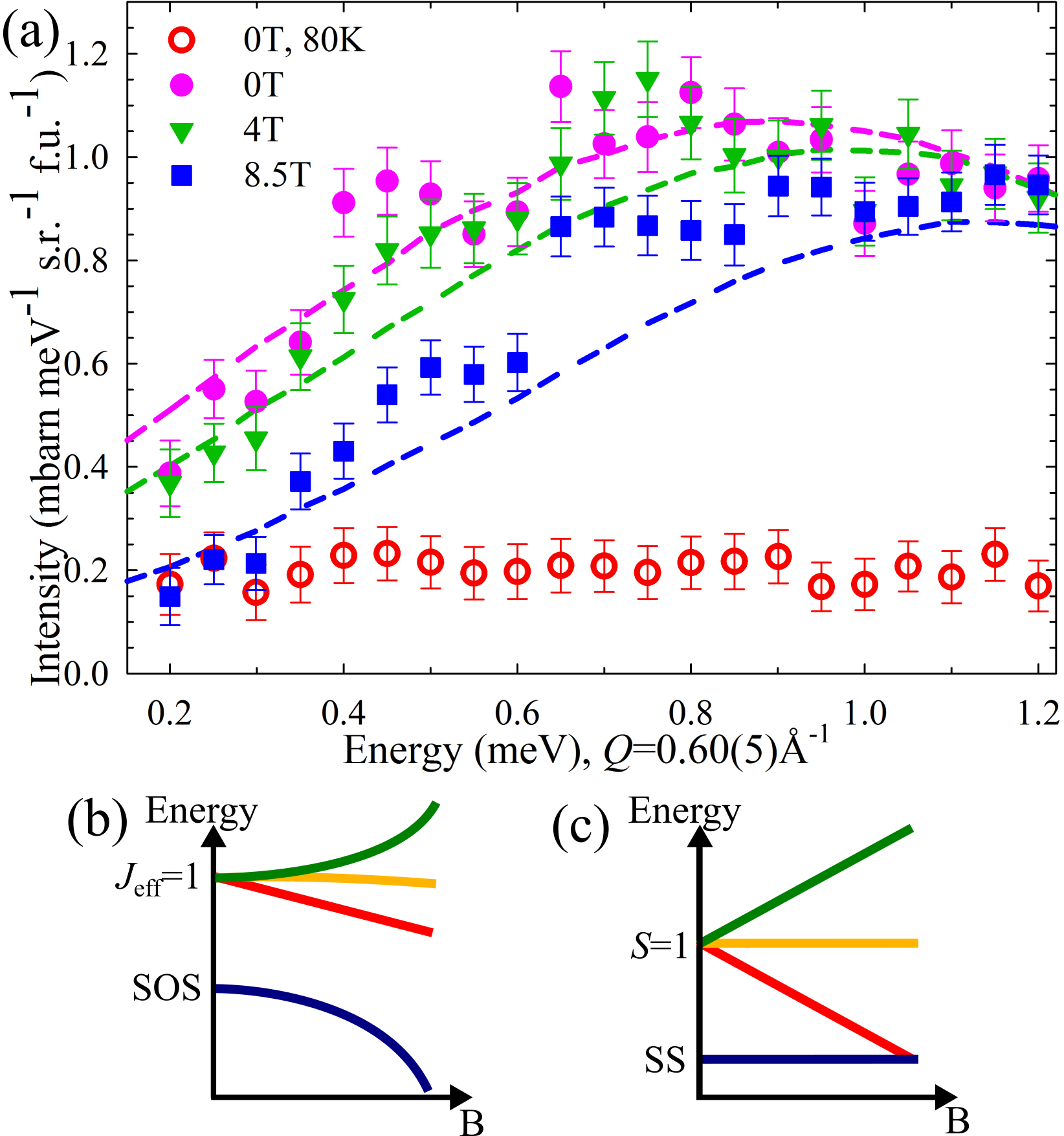}
\caption[]{(Color online) (a) Energy scan at the softening
wavevector $Q_S$ through the background subtracted data in
Fig.~\ref{fig:focus}(a-c): filled circles, triangles and squares
denote fields of 0, 4 and 8.5~T, respectively. Dashed lines are the model calculations shown in
Fig.~\ref{fig:focus}(d-f). Data at 0~T in the
paramagnetic phase at high temperatures (80~K) is also included
(red open circles). b) Schematic of the
ground and excited states' energies as a function of applied field
in the case of a SOS and (c) a spin-singlet system.}
\label{fig:focus2}
\end{center}
\end{figure}
%%%%%%%%%%%%%%%%%%%%%%%%%%%%%%%%%%%%%%%%%%%%%%%%%%%%%%%%%%%%%

The magnetic field dependence of the excitations was measured on
the same powder sample using the FOCUS time-of-flight spectrometer
at the Swiss Spallation Neutron Source SINQ (PSI) 
with the sample placed inside a vertical 9~T cryomagnet.
The obtained magnetic INS signal is plotted in
Figs.~\ref{fig:focus}(a-c). By comparing the data at different
fields it is apparent that the intense V-shaped magnetic signal
near $Q_S$ shifts upwards upon increasing field. This trend is directly seen
in the energy scan in Fig.~\ref{fig:focus2}(a) by comparing the
data at 0, 4 and 8.5~T, spectral weight moves to higher energies
upon increasing field.

%\section*{Analysis}
Below we compare quantitatively the dispersive features of the
excitation spectrum and the observed magnetic field dependence of
the low-energy scattering with a model of spin-orbital triplons of
a SOS ground state. In zero field the triplon dispersion derived
in the harmonic approximation using pseudo-boson operators
\cite{supplemental} or alternative methods \cite{ish} is
\begin{equation}
\hbar\omega(\bm{k})=\lambda\sqrt{1+\frac{4J(\bm{k})}{\lambda}},
\label{eq:omega}
\end{equation}
where $J(\bm{k})=J_2\sum_{\bm{A}}\cos (\bm{k}\cdot \bm{A})$ is the
Fourier transform of the exchange couplings and $\bm{A}$ runs over
all NN vectors of an FCC sublattice. The local singlet-triplet gap
$\lambda$ is determined by the crystal field strength
parameterized (using standard convention \cite{abrablean}) by the
single parameter $B_4<0$, and the atomic spin-orbit coupling
$\lambda_0<0$. Within a minimal ($J_2$,$B_4$,$\lambda_0$) model we
calculate the powder-averaged INS spectrum including the
triplons' dynamical structure factor (for details see
\cite{supplemental}) and compare systematically with scans through
the INS data as shown in Figs.~\ref{fig:merlin}(c) and
\ref{fig:focus2}(a) (magenta filled symbols). In addition, we
require the model parameters to reproduce optical data: the sharp
4.46~meV absorption in THz spectroscopy
\cite{laurita,mittelstadt}, identified with the triplon energy
(\ref{eq:omega}) at the zone center,
$\hbar\omega(\bm{0})=\sqrt{\lambda(\lambda+48J_2)}$, and the sharp
optical absorption at $\hbar\Omega=262(2)$~meV, attributed to the
transition from the ground state to the lowest level of the upper
orbital triplet \cite{note2}. Using those multiple constraints the
best fit parametrization is obtained for $J_2=0.136(6)$~meV,
$B_4=-2.45(6)$~meV and $\lambda_0=-12.1(1)$~meV, which give
$\lambda=2.28(6)$~meV \cite{note3}. This parametrization
reproduces (by construction) the energies of both optical
transitions and the $J_2$ value is comparable to that deduced from
Curie-Weiss fits of the high-temperature
susceptibility~\cite{laurita} and estimated from
density-functional calculations~\cite{sarkar}. The $B_4$ and
$\lambda_0$ values are comparable with $-2.58$ and $-10.0(9)$~meV,
respectively, found for \fep ions in FeCr$_2$S$_4$~\cite{feiner}.

The INS spectrum for the fitted parameter values is shown in
Fig.~\ref{fig:merlin}(b), where we have also included an
intrinsic linewidth broadening $\sim$1~meV, a possible
significance of this broadening will be discussed later. The
parametrization by the minimal model captures well the key
features of the INS data with clear V-shaped dispersions and mode
softenings near 0.6 and 1.4~\AA$^{-1}$, identified with scattering
emanating near the reciprocal lattice positions (100) and (211),
respectively. A corresponding calculation performed for the data measured on FOCUS at zero
field is shown in Fig.~\ref{fig:focus}(d) and this also
compares well with the data in panel (a). Energy scans near the
softening wavevector are in good agreement between the data and
model [see Fig.~\ref{fig:focus2}(a), magenta filled symbols/line].
Fig.~\ref{fig:merlin}(c) shows also the limitations of the present
model. The energy scan shown cuts across the low-$Q$ dispersion
and the model (solid line) reproduces well the observed peak
position. However, the linewidth is broader than expected based on
resolution effects alone (horizontal bar) and there is
considerable additional continuum scattering intensity at higher
energies above 3~meV, which we attribute to multi-triplon
scattering events, not included in the present model.

With the model parameters kept fixed by the fits to zero-field
data, we now calculate the expected behavior in an external
magnetic field, which contributes additional terms to the
single-ion Hamiltonian; $\mu_B \bm{B} \cdot (\bm{L}+2\bm{S}) +
12J_2\bm{S} \cdot \langle \bm{S} \rangle$. The first term is the
Zeeman energy in field and the second term includes the effect of
the exchange interactions, treated in a mean-field approximation
\cite{chen_prb}. $\langle \bm{S} \rangle$ is the field-induced
spin polarization of the ground state, i.e. $\langle \bm{S}
\rangle=\langle\psi_0|\bm{S}|\psi_0\rangle$, where $\psi_0$ is the
ground state wavefunction of the single-site Hamiltonian. Solving
for $\langle \bm{S} \rangle$ self-consistently we find the
wavefunctions $\psi_{1,2,3}$ and energies $\lambda_{1,2,3}$ for a
general field direction, determine the triplon dispersion
relations and neutron structure factor, then average the spectrum
over a spherically uniform distribution of powder grains (see
\cite{supplemental} for details). The model calculations are
compared with the measured INS data in Fig.~\ref{fig:focus},
panels (e-f) with (b-c) at 4 and 8.5~T; the model captures the
apparent upwards shift of the scattering intensity upon increasing
field. This is even more clearly seen in the energy scans in
Fig.~\ref{fig:focus2}(a), the model calculation (dashed lines)
reproduce well the observed shift of spectral weight to higher
energies upon increasing field with no adjustable parameters once
the overall intensity scale factor is fixed by the comparison in
zero field.

It is insightful to compare the spin-orbital triplons of a SOS
ground state discussed here with triplons of a spin-singlet (SS)
ground state with completely quenched orbital degree of freedom,
as found for example in quantum dimerized antiferromagnets like
TlCuCl$_3$ \cite{ruegg}. For the latter, a magnetic field Zeeman
splits the triplet into $S_z=-1,0,+1$ states, with a linear
reduction in the gap to the $S_z=-1$ state, as shown schematically
in Fig.~\ref{fig:focus2}(c). At a critical field level crossing
with the ground state occurs and magnetic order ensues via
condensation of triplons. One might wonder how the behavior of
spin-orbital triplons can be any different; the triplons now have
an effective angular momentum $J_{\rm eff}=1$ (as opposed to $S=1$
in the SS case). A low applied magnetic field Zeeman splits the
triplet into $J_{\rm{eff},z}=-1,0,+1$ states \cite{low,laurita},
however at higher fields terms quadratic and higher in $B$ prevail
\cite{ish,note1} and allow mixing between the SOS and the
$J_{\rm{eff},z}=0$ triplet mode. This enables the ground state to
{\em reduce} its energy in applied field by acquiring a finite
polarization along the field direction, see
Fig.~\ref{fig:focus2}(b), thus avoiding magnetic order via level
crossing with the triplet states.

We now relate our results to the generic phase diagram in
Fig.~\ref{fig:structure}, describing the transition from SOS
to magnetic/orbital order upon increasing $x = J_2/\lambda$. Using
the parameters obtained from fitting the INS data yields
$x\simeq0.060$, marginally close to the proposed critical value
$x_c=1/16$. For such close proximity to criticality one might
expect manifestations of enhanced quantum fluctuations associated
with the critical point. For $x \lesssim x_c$ in addition to sharp
triplon excitations one would also expect multi-triplon continua
at higher energies, with enhanced spectral weight and decreasing
gap as $x \nearrow x_c$, with the triplon dispersions becoming
lower boundaries of a critical continuum of excitations precisely
at the quantum critical point at $x_c$. Effects associated with such continuum scattering and/or
broadening of sharp modes may be at least partly responsible for
the extra scattering intensity and broadening effects observed in
the INS data in Fig.~\ref{fig:merlin}(c), 
we hope our results will stimulate further
theoretical modelling of such effects close to spin-orbital
quantum criticality.

One may ask if other materials may exhibit related physics. We
note that a high-spin $d^4$ ion (e.g. Mn$^{3+}$) in an octahedral
(weak) cubic crystal field displays the same single-ion physics
({\em electron} analogue) as \fep in \fesc, i.e. spin $S=2$ and
{\em two}-fold orbital degeneracy, where the spin-orbit coupling
(now $\lambda_0>0$) stabilizes the SOS ground state in
(\ref{eq:gsideal}) with a $J_{\rm eff}=1$ excited triplet. Similar
singlet-triplet physics, but with a singlet ground state distinct
from (\ref{eq:gsideal}), originating from $S=1$ and {\em
three}-fold orbital degeneracy, is expected for low-spin $d^4$
ions (e.g. Ru$^{4+}$) in strong octahedral crystal field
\cite{excitonic_magnetism,ruthenate} and $d^8$ ions (e.g.
Ni$^{2+}$) in tetrahedral field \cite{note4}. If such ions can resist JT
distortions, they are candidates to display correlated spin-orbit
states under inter-site exchange, potentially in a different part
of the phase diagram in Fig.~\ref{fig:structure}.

%\section*{Conclusions}
To summarize, we have reported powder INS measurements of the full
bandwidth of magnetic excitations in the spinel \fesc and have %AFM
found that that the key dispersive features can be well described
by spin-orbital triplons of a near-critical SOS state. In high
applied magnetic field we have observed a shift of spectral
weight to higher energies, giving support to the theoretical
proposal \cite{chen_prb} that applied fields further stabilize the
SOS state by moving the system away from the quantum critical
point, this is a direct consequence of the entangled spin-orbital nature
of the ground and excited triplet states.

\begin{acknowledgments}
This work was partially supported by the EPSRC (U.K) under Grants No. EP/H014934/1 and EP/M020517/1 as well as the SNF SCOPES project
IZ73Z0\_152734/1, the Marie Curie FP7 COFUND PSI Fellowship
program, Swiss National Science Foundation, Sinergia Network Mott Physics Beyond the Heisenberg Model, the ERC Grant Hyper Quantum Criticality (HyperQC)
 and Transregional Research Collaboration TRR 80 (Augsburg,
Munich, Stuttgart). This work is partially based on experiments performed 
at the Swiss spallation neutron source SINQ, Paul Scherrer Institute, Villigen, Switzerland. 
In accordance with the EPSRC policy framework on research data, access to the data will be made available from Ref. \cite{EPSRCdata}
\end{acknowledgments}

{\em Note added.} As this work was being completed
Ref.~\cite{plumb} appeared, reporting evidence for marginal
magnetic order in samples synthesized using a different protocol,
suggesting an extreme sensitivity to the synthesis route. Broadly speaking, there are three main physical factors that could lead to such a discrepancy; vacancies, site disorder, and off-stoichiometry, all of which are discussed in the Supplemental Material \cite{supplemental}. We conclude that off-stoichiometry can lead to magnetic order with a transition temperature of a few K.
Our results highlighting that magnetic fields favor the SOS state
suggest a very interesting possibility that fields applied onto an
ordered sample, potentially along a particular direction in a
single crystal, may drive it towards the SOS state and thus reach
the long-searched-for quantum critical point.

\bibliography{FeSc2S4_Bibliography_arxiv}

\begin{thebibliography}{34}
\bibitem [{not({\natexlab{a}})}]{note2}%
  \BibitemOpen
  \href@noop {} {}\bibinfo {note} {{A}s discussed in \cite{wittekoek,laurita},
  the transition from the ground state to the lowest level of the orbital
  triplet occurs at $\hbar\Omega=\hbar\Omega_0 - E_{\rm JT}$, where $E_{\rm
  JT}$ is an energy shift due to the coupling to Jahn-Teller phonons. We have
  used the estimate $E_{\rm JT}\approx7$~meV \cite{laurita} to deduce the
  energy of the purely electronic transition $\hbar\Omega_0$ and this was then
  used in the data parameterization to constrain $B_4$ and
  $\lambda_0$.}\BibitemShut {Stop}%
\bibitem [{not({\natexlab{b}})}]{note3}%
  \BibitemOpen
  \href@noop {} {}\bibinfo {note} {{N}ote that the usually-assumed lowest order
  approximation $\lambda\approx6\lambda_0^2/(120|B_4|)$ \cite{low} would
  predict a value $\sim$30\% higher than that obtained from directly
  calculating the energy levels of the full single-ion
  Hamiltonian.}\BibitemShut {Stop}%
\bibitem [{not({\natexlab{c}})}]{note1}%
  \BibitemOpen
  \href@noop {} {}\bibinfo {note} {{T}he $g$-factor characterizing the linear
  splitting is predicted \cite{ish} to be wavevector-dependent and vary as the
  square of the zero-field triplon energy in (\ref{eq:omega}), therefore to
  become negligibly small near the magnetic softening wavevector, where the
  regime of linear splitting of the triplet modes is practically unobservable
  as quadratic and higher order terms in $B$ dominate.}\BibitemShut {Stop}%
\end{thebibliography}

\section{Supplemental Material}
\makeatletter
\renewcommand{\thefigure}{S\@arabic\c@figure}
\renewcommand{\theequation}{S\@arabic\c@equation}
\renewcommand{\thetable}{S\@arabic\c@table}
\makeatother
\setcounter{figure}{0}
\setcounter{equation}{0}

Here we outline 1) the derivation of the spin-orbital
wavefunctions for a single Fe$^{2+}$ ion in a tetrahedral cubic
crystal field including spin-orbit coupling, 2) the description of
the lowest singlet-triplet transition in terms of spin-orbital
triplon operators, 3) the analytic derivation of the triplon
dispersions in the presence of magnetic exchange interactions and
the relevant matrix elements for neutron scattering, 4) the
derivation of single-ion states in the presence of an external
magnetic field and exchange via a mean-field approach, 5) the
non-magnetic background subtraction procedure for the INS data via
the principle of detailed balance, 6) the derivation of the
neutron cross-section for triplon scattering and spherical
averaging to compare with powder INS data, and 7) details on the
sample preparation for the \fesc powder used in the INS
experiments.

\section{S1. Single Ion Hamiltonian}
\label{fesc:singleion}

This section outlines the derivation of the spin-orbital
wavefunctions for a singe \fep ($3d^6$) ion in the (weak) crystal
field environment appropriate for \fesc. The Hamiltonian is
\begin{equation}
\label{eq:Hfirstap} \mathcal{H}=\mathcal{H}_{\rm
cf}+\mathcal{H}_{\rm SO}+\mathcal{H}_\text{Zeeman},
\end{equation}
where the three terms are the crystal field, spin-orbit and
external magnetic field contributions, respectively. The
crystal-field term can be expressed via the equivalent operator
method in terms of Stevens operators of the orbital angular
momentum $\bm{L}$. The allowed terms are constrained by the local
site symmetry and for a cubic environment $\mathcal{H}_{\rm cf}$
is of the form \cite{abrablean}
\begin{equation} %\label{eq:Hcfap}
\mathcal{H}_{\rm cf}=B_4(O_4^0+5O_4^4), \nonumber
\end{equation}
where $O_4^0$ and $O_4^4$ are Stevens operators (tabulated in
\cite{hutch}) and $B_4$ is a constant that characterizes the
strength of the crystal field ($B_4<0$ for a $d^6$ ion in
tetrahedral coordination). In expanded form the crystal-field
Hamiltonian reads
\begin{equation}
\begin{array}{ll}
\mathcal{H}_{\rm cf}=B_4([35L_z^4-30L(L+1)L_z^2+25L_z^2\\
~~-6L(L+1)+3L^2(L+1)^2]+\frac{5}{2}[L_+^4+L_-^4]), &
\end{array}
\label{eq:Hcfexap}
\end{equation}
where the Cartesian $x,y,z$ axes are chosen along the cubic axes
of the unit cell.

The second term in (\ref{eq:Hfirstap}) is the atomic spin-orbit
interaction,
\begin{equation}
\mathcal{H}_{\rm SO}=\lambda_0 \bm{L} \cdot \bm{S},
\label{eq:Hso}
\end{equation}
with $\lambda_0<0$ for a $d^6$ ion (hole-like). For calculation
purposes it is helpful to expand the dot product as
\begin{equation} %\label{eq:LS}
\bm{L} \cdot \bm{S}=L_zS_z+\frac{1}{2}\left(L_+S_-+ L_-S_+\right),
\nonumber
\end{equation}
where the $\pm$ ladder operators are the standard ones, i.e.
$L_{\pm}|L,M_L\rangle=\sqrt{(L\mp M_L)(L\pm M_L+1)}|L,M_L\pm
1\rangle $ and similar for $S_{\pm}$.
\begin{figure}[htbp]
\includegraphics[width=0.48\textwidth]{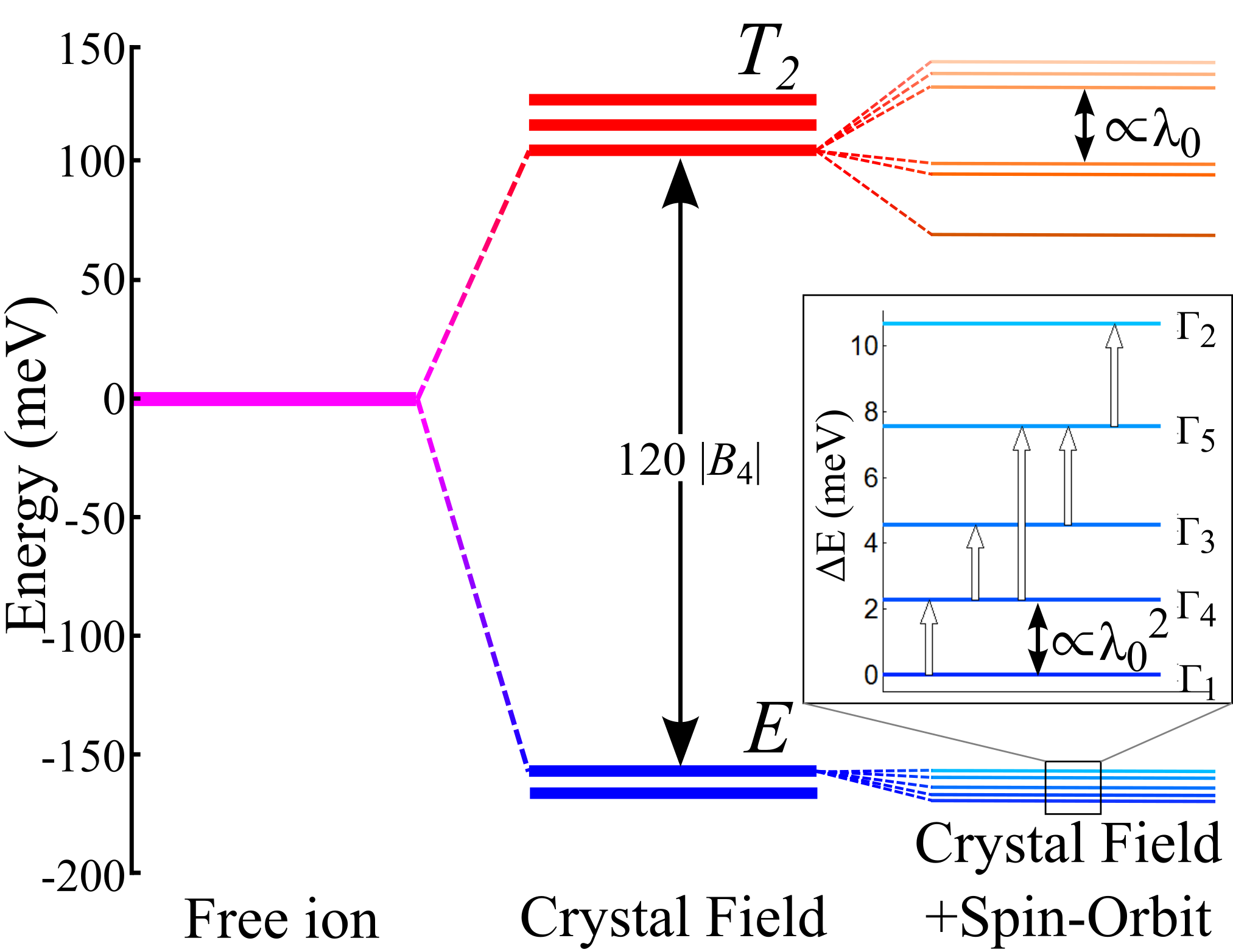}
\caption[]{(Color online) Splitting of Fe$^{2+}$ free ion orbital
levels due to crystal field and spin-orbit coupling. Inset:
Allowed transitions via neutron scattering for the 5 states
derived from the $E$-doublet where side labels $\Gamma_{1-5}$
indicate the corresponding irreducible representations (from
\cite{test}).}\label{fig:cf}
\end{figure}

The Fe$^{2+}$ ($3d^6$) ions are in an $S=2$, $L=2$ configuration.
We use the $|M_L, M_S \rangle$ states as basis to describe the
wavefunctions, where $M_L$ and $M_S$ are the projections of the
$\bm{L}$ and $\bm{S}$ operators onto the quantization axis $z$,
each takes values of $-2 \dots 2$. In this basis all operators are
represented by $25\times25$ matrices and diagonalization of the
Hamiltonian (\ref{eq:Hfirstap}) obtains the spectrum of states
shown in Fig.~\ref{fig:cf}. The cubic crystal field splits the
5-fold degenerate $L=2$ orbital states into a lower $E$-doublet
and upper $T_2$-triplet above a gap $\Delta_{\rm cf}=-120B_4>0$
(same level splitting, symmetry of wavefunctions and order of
levels as for a single $d$-electron with orbital quantum number
$l=2$). The spin-orbit coupling further splits those levels. At
lowest order in $\lambda_0$, the upper $T_2$ manifold is split
into three levels with energies $\Delta_{\rm cf}-3\lambda_0$,
$\Delta_{\rm cf}-\lambda_0$ and $\Delta_{\rm cf}+2\lambda_0$, and
the $E$ manifold is split into 5 equidistant levels separated by
$6\lambda_0^2/\Delta_{\rm cf}$ \cite{low}. The last column in
Fig.~\ref{fig:cf} indicates those lowest five levels and their
irreducible representations, the ground state is a singlet and the
first excited state a triplet. Transitions between states probed
via neutron scattering are determined by the matrix element
\begin{equation} \label{eq:int}
\langle f| \bm{L} + 2 \bm{S}|i \rangle, %\nonumber
\end{equation}
where $|i\rangle$ and $|f\rangle$ denote the initial and final
states, respectively. The symmetry-allowed transitions between
levels originating from the lower $E$-doublet are indicated by
thick white arrows in Fig.~\ref{fig:cf} last column.

Note that the three degenerate states of the first excited triplet
can be described by an effective angular momentum $J
_\text{eff}=1$ with $\psi_{1,2,3}$ identified as eigenstates of
$J_{\text{eff},z}$ with eigenvalues of $-1,0,+1$ respectively. To
obtain the explicit wavefunctions of those states (listed in
Table~\ref{tb:wavefunctions}) we solve for the eigenstates in the
presence of an infinitesimally small applied magnetic field
$\bm{B}\parallel(001)$ (to be discussed later) and then choose
appropriate relative signs in front of the obtained wavefunctions
such that the they satisfy the operator algebra for the total
angular momentum $\bm{J}=\bm{L}+\bm{S}\equiv\alpha \bm{J}_{\rm
eff}$. Explicitly, the matrix representations of the operators
$J_z$ and $J_+$ in the basis of $\psi_{1,2,3}$ states are found to
be
\begin{equation}
J_z = \alpha \left( \begin{array}{ccc}
-1 & 0 & 0 \\
 0 & 0 & 0  \\
 0 & 0 & 1
\end{array} \right),
~~~ J_+ = \alpha \left( \begin{array}{ccc}
0 & 0 & 0 \\
\sqrt{2} & 0 & 0 \\
0 & \sqrt{2} & 0
\end{array} \right) \nonumber
%\label{eq:jeffective3}
\end{equation}
where the projection factor is $\alpha \approx 0.44$ for the $B_4$
and $\lambda_0$ values used in the analysis ($\alpha \rightarrow
1/2$ as $|B_4| \rightarrow \infty$).

\begin{table}[htbp]
\begin{center}
\begin{tabular}{|l|c|c|c|c|}
\hline
$|M_L, M_S \rangle$  & $\psi_0$ & $\psi _1 $ & $\psi _2 $ & $\psi _3$ \\
\hline
$|2,2 \rangle$ & $0.399$  &  -  &  $-0.571$  &  -  \\
\hline
$|2,1 \rangle$ & -  &  $-0.371$  &  -  &  -  \\
\hline
$|2,-1 \rangle$ & -  &  -  &  -  &  $0.326$  \\
\hline
$|2,-2 \rangle$ &  $0.297$  &  -  &  $0.416$  &  -  \\
\hline
$|1,2 \rangle$  &  -  &  $-0.032$  &  -  &  -  \\
\hline
$|1,0 \rangle$  &  -  &  -  &  -  &  $0.135$  \\
\hline
$|1,-1 \rangle$ & $0.102$  &  -  &  $0.032$  &  -  \\
\hline
$|1,-2 \rangle$ & -  &  $-0.078$  &  -  &  -  \\
\hline
$|0,1 \rangle$  & -  &  -  &  -  &  $0.855$  \\
\hline
$|0,0 \rangle$  &  $0.696$  &  -  &  -  &  -  \\
\hline
$|0,-1 \rangle$ & -  &  $-0.855$  &  -  &  -  \\
\hline
$|$-$1,2 \rangle$ &  -  &  -  &  -  &  $0.078$  \\
\hline
$|-1,1 \rangle$  & $0.102$  &  -  &  $-0.032$  &  -  \\
\hline
$|-1,0 \rangle$  &  -  &  $-0.135$  &  -  &  -  \\
\hline
$|-1,-2 \rangle$ &  -  &  -  &  -  &  $0.032$  \\
\hline
$|-2,2 \rangle$  & $0.297$  &  -  &  $-0.416$  &  -  \\
\hline
$|-2,1 \rangle$  &  -  &  $-0.326$  &  -  &  -  \\
\hline
$|-2,-1 \rangle$ & -  &  -  &  -  &  $0.372$  \\
\hline
$|-2,-2 \rangle$ & $0.399$  &  -  &  $0.572$  &  -  \\
\hline
\end{tabular}
\end{center}
\caption{The wavefunctions $\psi_{0-3}$ of the four lowest energy
single-ion states for $\lambda_0=-12.1$~meV and $B_4=-2.45$~meV
($B$=0) expressed in the 25-element basis of $| M_L, M_S \rangle$
states. For those parameters the energy gap for the transition
$\psi_0\rightarrow\psi_{1,2,3}$ is $\lambda=2.28$~meV. Note that
the lowest order approximation
$\lambda\approx6\lambda_0^2/(120|B_4|)$ is not applicable for the
present case as it would predict a $\sim$30\% higher value than
obtained from directly calculating the energy levels of the full
single-ion Hamiltonian in (\ref{eq:Hfirstap}).}
\label{tb:wavefunctions}
\end{table}

We have explicitly verified that the wavefunctions obtained agree
with previous studies \cite{test, abrablean} of \fep ions in cubic
crystal-field environments. Furthermore, we have verified that the
spin-orbit coupling only mixes states belonging to the same
irreducible representation, as expected from symmetry
considerations. For example, the ground state wavefunction in
Table~\ref{tb:wavefunctions} can be written as
\begin{equation}
\begin{array}{ll}
\psi_0 \approx & 0.984 \times \Big[\frac{1}{\sqrt{2}}|3z^2-r^2\rangle|S^z=0\rangle   \\
& ~ +\frac{1}{2}|x^2-y^2\rangle(|S_z=-2\rangle+|S_z=2\rangle) \Big] \\
& - 0.178 \times \frac{1}{\sqrt{3}}\Big[ \frac{1}{\sqrt{2}i}|xy \rangle(|S_z=2\rangle-|S_z=-2\rangle) \\
& ~+ \frac{1}{\sqrt{2}i}|yz \rangle\rangle(-|S_z=1\rangle-|S_z=-1\rangle) \\
& ~+ \frac{1}{\sqrt{2}} |zx \rangle(|S_z=-2\rangle+|S_z=2\rangle)
\Big], \label{eq:wvfnct_nice}
\end{array}
\end{equation}
where the usual notation for $d$-orbitals has been used. In the
above expansion the first term is the ``ideal'' SOS state in (1)
(obtained in the limit $|B_4| \rightarrow \infty$). The second
term in (\ref{eq:wvfnct_nice}) is a singlet state originating from
the $T_2$ level, mixed in by the spin-orbit coupling.

In a finite magnetic field the single-ion Hamiltonian
(\ref{eq:Hfirstap}) acquires a Zeeman term,
\begin{equation} \label{eq:HZeeap}
\mathcal{H}_\text{Zeeman}=\mu_B{\bm B} \cdot (\bm L + g_S \bm S),
\end{equation}
where we assume $g_S=2$ for spin. The magnetic field dependence of
the energy levels of the four lowest states is schematically
illustrated in Fig.~4(b), the three lowest excited states
$\psi_{1,2,3}$ have now distinct energy gaps $\lambda_{1,2,3}$
above the ground state. In the limit of small applied field the
behavior is isotropic, independent of the applied field direction,
and the splitting of the triplet states can be described by an
effective Zeeman term $\mathcal{H}_\text{Zeeman, eff}= g\mu_B{\bm
B} \cdot {\bm J}_{\rm eff}$. For the $B_4$ and $\lambda_0$ values
used here the $g$-factor is obtained as $g\approx0.94$. For
moderate magnitude applied fields (when the Zeeman energy is
comparable to the zero-field gap $\lambda$) the splitting of the
excited triplet is non-linear, cannot be described in terms of the
simplified $J_{\rm eff}=1$ states, and furthermore is strongly
dependent on the applied field direction with respect to the cubic
axes, so in the general case we determine the wavefunctions
$\psi_{0-4}$ of the four lowest states and the gaps
$\lambda_{1-3}$ via a direct diagonalization of the full
single-ion Hamiltonian in (\ref{eq:Hfirstap}).

%%%%%%%%%%%%%%%%%%%%%%%%%%%%%%%%%%%%%%%%%%%%%%%%%%%%%%%%%%%%%
\section{S2. Pseudo-Boson Operators}
\label{fesc2s4:psbote}
%%%%%%%%%%%%%%%%%%%%%%%%%%%%%%%%%%%%%%%%%%%%%%%%%%%%%%%%%%%%%

In this section we introduce pseudo-boson operators to describe
the transitions from the $\psi_0$ ground state singlet to the
$\psi_{1,2,3}$ excited triplet states to have a physical basis to
describe the magnetic dynamics. At very low temperatures only the
$\psi_0$ ground state is thermally populated and the only
symmetry-allowed transitions in neutron scattering are to the
$\psi_{1,2,3}$ states. So to capture the magnetic dynamics
observable by neutron scattering it is sufficient to consider the
restricted set of those fours basis states and construct matrix
representations of all operators in this restricted basis, i.e.
for a general operator $\widehat{O}$ this would be
\begin{equation} \label{eq:mat}
\widehat{O} = \left[ \begin{array}{ccc}
\langle\psi_0|\widehat{O}|\psi_0\rangle & \langle\psi_0|\widehat{O}|\psi_1\rangle & \ldots \\
\langle\psi_1|\widehat{O}|\psi_0\rangle & \langle\psi_1|\widehat{O}|\psi_1\rangle & \ldots \\
\vdots & \vdots & \ddots \\
\end{array} \right].
\end{equation}
An alternative description of the restricted set of basis states
is in terms of occupation numbers of three types of pseudo-bosons
\cite{grover}, where the ground state $\psi_0$ is interpreted as
the `vacuum', the excited $\psi_1$ state corresponds to having one
$a$- type boson present, the $\psi_2$ has one $b$-type boson, and
so on. Explicitly, the creation operators for the three types of
bosons are defined as
\begin{eqnarray}
a^{\dagger}\psi_0 & = & \psi_1,\nonumber \\
b^{\dagger}\psi_0 & = & \psi_2, \nonumber \\
c^{\dagger}\psi_0 & = & \psi_3, %\nonumber
\label{eq:abc}
\end{eqnarray}
where the annihilation operators are obtained by Hermitian
conjugation as $ a\psi_1 =  \psi_0$ and so on. The pseudo-boson
operators have a trivial matrix representation in terms of the
four-basis states $\psi_{0-3}$, i.e. the creation operator for
$a$-type bosons is represented as
\begin{equation}
\label{eq:mat2} a^{\dagger} = \left[
\begin{array}{cccc}
0 & 0 & 0 & 0 \\
1 & 0 & 0 & 0 \\
0 & 0 & 0 & 0 \\
0 & 0 & 0 & 0 \\
\end{array} \right].
\end{equation}

By comparing (\ref{eq:mat}) and representations of the type shown
in (\ref{eq:mat2}) it is clear that the $4\times4$ matrix
representation of a general operator may be equivalently expressed
as a sum of linear and bilinear terms in the boson
creation/annihilation operators, and the identity operator.
Therefore, once the wavefunctions $\psi_{0-3}$ are known
explicitly, then the $4\times4$ matrix representation of all
relevant spin and orbital operator components such as $S_z,
L_z,\dots$ can be constructed, and those can then be expanded in
terms of boson operators.

\section{S3. Dispersion of Triplons}
\label{fesc2s4:disptrip}

In this section we outline the derivation of the dispersion
relations of magnetic excitations in the harmonic approximation
using the pseudo-boson operators defined in the previous section.
In the presence of magnetic exchange interactions between \fep
sites the local singlet-triplet transitions acquire a
dispersion, i.e. the pseudo-bosons become delocalized by hopping
across lattice sites. This leads to coherently-propagating
excitations, so-called `spin-orbital triplons' due to the mixed
spin-orbital character and the three-fold degeneracy (in zero
field).

We assume in a first approximation that the global ground state is
the same as in the non-interacting case, given by the product of
the $\psi_0(\bm{r})$ single-ion states at all sites $\bm{r}$ in
the lattice, and we focus on the effects of the exchange
interactions on the singlet-triplet transition.
Considering a minimal model with a Heisenberg antiferromagnetic
exchange interaction $J_2>0$, the Hamiltonian for each of the two
magnetically-decoupled FCC sublattices reads
\begin{equation} \label{eq:ex}
\mathcal{H}_{\text{ex}}=\sum_{\langle ij \rangle}^{}J_{2}\bm{S_i}
\cdot \bm{S_j},
\end{equation}
where the sum extends over all $ij$ NN pairs of sites, with each
pair counted once. In expanded form this reads
\begin{equation}
\begin{array}{l l}
\mathcal{H}_{\text{ex}} & =  J_2 \sum_{\langle ij \rangle }^{}(S_z(\bm{r}_i)S_z(\bm{r}_j) + \\
 & \frac{1}{2}[S_+(\bm{r}_i)S_-(\bm{r}_j)+S_-(\bm{r}_i)S_+(\bm{r}_j)]), \\
\end{array}
 \label{eq:spinexap}
\end{equation}
where $S_z(\bm{r}_i)$ is the $z$-component of the spin operator at
site $\bm{r}_i$ on the lattice, and so on. The goal is to convert
the exchange Hamiltonian from spin operators to boson
creation/annihilation operators. The spin operator components are
found to have the following expansion in terms of boson operators
\begin{eqnarray}
S_+(\bm{r})& = & f_1 a(\bm{r})+f_2 c^{\dagger}(\bm{r})+\dots ,\nonumber \\
S_-(\bm{r})& = & f_1 a^{\dagger}(\bm{r})+f_2 c(\bm{r})+\dots,\nonumber \\
S_z(\bm{r})& = & f_3 \left[b(\bm{r})+
b^{\dagger}(\bm{r})\right]+\dots, \label{eq:szed}
\end{eqnarray}
where only the leading (linear) terms are given. The above
expansion is valid in the case of an applied magnetic field along
one of the cubic axes, labelled $z$ (the case for a general field
orientation will be discussed later). The pre-factors in the
expansion, $f_1=\langle\psi_0| S_+ |\psi_1 \rangle$,
$f_2=\langle\psi_3| S_+ |\psi_0 \rangle$ and $f_3=\langle\psi_0|
S_z |\psi_2 \rangle$, are matrix elements that depend on the
wavefunction content of the lowest four states, $\psi_{0-3}$,
which in turn depend on $B_4$, $\lambda_0$ and the magnetic field
strength $B$. In (\ref{eq:szed}) we have explicitly included the
position dependence of the operators, i.e. $a^{\dagger}(\bm{r})$
creates an $a$-type boson at site $\bm{r}$ and so on.

Substituting (\ref{eq:szed}) into (\ref{eq:spinexap}) gives the
spin Hamiltonian as a bilinear form of boson operators. To allow
this to be diagonalized to find the normal modes we introduce the
Fourier-transformed operators defined by
\begin{equation}
a(\bm{r})  =
\frac{1}{\sqrt{N}}\sum_{{\bm{k}}}^{}a_{{\bm{k}}}e^{-i {\bm{k}}
\cdot \bm{r}}, %\nonumber
\end{equation}
with similar expressions for $b(\bm{r})$ and $c(\bm{r})$. Here $N$
is the number of sites in an FCC sublattice. The exchange
Hamiltonian expanded up to quadratic order in the bosons reads
\begin{equation} \label{eq:white}
\mathcal{H}_{\rm ex}=\sum_{\bm{k}} X^{\dagger} H X + E_0,
\end{equation}
where $E_0$ is a constant, the sum extends all wavevectors
$\bm{k}$ in the Brillouin zone of the FCC sublattice and the
$\bm{k}$ dependence of the operator matrix $X$ and of the
(Hermitian) Hamiltonian matrix $H$ is implicit. The operator
matrix $X^{\dagger}$ is the row vector
\begin{equation} \label{eq:Xdag}
X^{\dagger}=[a_{{\bm{k}}}^{\dagger}~c_{{\bm{k}}}^{\dagger}~b_{{\bm{k}}}^{\dagger}~a_{-{{\bm{k}}}}~c_{-{{\bm{k}}}}~
b_{-{{\bm{k}}}}],
\end{equation}
and $X$ is its adjoint column vector. The Hamiltonian 6$\times$6
matrix $H$ has the block form
\begin{equation} \label{eq:par}
H= \left [ \begin{array}{cc}
A_{\bm{k}} & B_{\bm{k}}\\
B_{\bm{k}} & A_{\bm{k}}\\
\end{array} \right],
\end{equation}
where
\begin{equation} %\label{eq:Amat}
A_{\bm{k}}= \frac{1}{2}\left [
\begin{array}{ccc}
\lambda_1 + \frac{f_1^2}{2}J(\bm{k}) & 0 & 0 \\
0 & \lambda_3 + \frac{f_2^2}{2}J(\bm{k}) & 0 \\
0 & 0 & \lambda_2 + f_3^2 J(\bm{k}) \\
\end{array}
\right],\nonumber
\end{equation}
\begin{equation} \label{eq:Bmat}
B_{\bm{k}}=\frac{J(\bm{k})}{2} \left[
\begin{array}{ccc}
0 & \frac{f_1 f_2}{2} & 0\\
\frac{f_1 f_2}{2} & 0 & 0\\
0 & 0 & f_3^2\\
\end{array} \right],
\end{equation}
and $J(\bm{k})$ is the Fourier transform of the exchange
interactions defined using the convention in \cite{ish} as
\begin{equation} %\label{eq:gammak}
J(\bm{k})=J_2\sum_{\bm{A}}\cos({\bm{k}} \cdot \bm{A}), \nonumber
\end{equation}
where the sum extends over all vectors $\bm{A}$ linking a
Fe$^{2+}$ ion to its 12 nearest neighbors on the same FCC
sublattice. $\lambda_{1,2,3}$ denote the energy cost of creating
an $a$-, $b$-, $c$-type boson, respectively, at a lattice site in
the absence of exchange interactions ($J_2=0$), with the three
levels being degenerate in zero field. We note that the order of
the operators in the row basis vector $X^{\dagger}$ listed in
(\ref{eq:Xdag}) was chosen such as to have a block form for the
matrix $B_{\bm{k}}$ in (\ref{eq:Bmat}).

Diagonalizing the Hamiltonian (\ref{eq:par}) using standard
methods \cite{white} leads to the following dispersion relations
for the triplons
\begin{eqnarray} %\label{eq:w1disp}
\hbar\omega_1({\bm{k}})&=&\left[\frac{f_1^2
J(\bm{k})}{4}+\frac{\lambda_1}{2}\right]\left[1-\phi_1+\sqrt{(1+\phi_1)^2-\xi_1^2}\right], \nonumber \\
%\label{eq:w3disp}
\hbar\omega_3({\bm{k}})&=&\left[\frac{f_2^2
J(\bm{k})}{4}+\frac{\lambda_3}{2}\right]\left[1-\phi_2+\sqrt{(1+\phi_2)^2-\xi_2^2}\right], \nonumber \\
\hbar\omega_2({\bm{k}})&=&\lambda_2 \sqrt{1+2\theta},
\label{eq:wdisp}
\end{eqnarray}
where
\begin{eqnarray}
\phi_1 & = & \frac{f_2^2 J(\bm{k})+2\lambda_3}{f_1^2 J(\bm{k})+2\lambda_1},\nonumber \\
\phi_2 & = & \frac{f_1^2 J(\bm{k})+2\lambda_1}{f_2^2 J(\bm{k})+2\lambda_3}, \nonumber \\
\xi_1  & = & \frac{2 f_1 f_2 J(\bm{k})}{f_1^2 J(\bm{k})+2\lambda_1}, \nonumber \\
\xi_2  & = & \frac{2 f_1 f_2 J(\bm{k})}{f_2^2 J(\bm{k})+2\lambda_3},\nonumber \\
\theta & = & \frac{f_3^2 J(\bm{k})}{\lambda_2}. \nonumber
\end{eqnarray} %\label{eq:phisandxis}
In zero field all three modes are degenerate and in the limit
$|B_4|\rightarrow \infty$, $|f_1|^2=|f_2|^2=4$ and $|f_3|^2=2$, so
the triplon dispersion becomes
\begin{equation}
\hbar\omega(\bm{k})=\lambda\sqrt{1+\frac{4J(\bm{k})}{\lambda}},
\end{equation}
in agreement with results deduced using a random-phase
approximation \cite{ish} and an earlier derivation \cite{chen_prl}
using a first order expansion in the exchange
$\hbar\omega(\bm{k})\simeq\lambda+2J(\bm{k})$.

In order to evaluate the matrix elements for the neutron
cross-section from triplons one also needs to know explicitly the
transformation matrix $\mathsf{Q}$ to the basis $Y$ of normal
operators where the Hamiltonian is diagonal. The transformation
matrix $\mathsf{Q}$ needs to satisfy the following three
conditions \cite{maestro}
\begin{eqnarray} %\label{eq:maestro1}
\mathsf{Q}\mathsf{g} \Lambda \mathsf{Q}^{-1} &=& \mathsf{g} H, \nonumber \\
\mathsf{Q}\mathsf{g}\mathsf{Q}^{\dagger}&=&\mathsf{g}, \nonumber \\
\mathsf{Q}^{\dagger} H \mathsf{Q}&=&\Lambda, \label{eq:maestro3}
\end{eqnarray}
where $\Lambda$ is the diagonal form of the Hamiltonian matrix and
$\mathsf{g}$ is the operator commutator matrix
\begin{equation} %\label{eq:gdef}
\mathsf{g}=[X^{\dagger},X]. \nonumber
\end{equation}
The normal operator basis $Y$ is related to the original operator
basis $X$ via
\begin{equation} \label{eq:xtoy}
X=\mathsf{Q}Y,
\end{equation}
where the row vector $Y^{\dagger}$ contains the normal boson
operators
\begin{equation}
Y^{\dagger}=[a'^{\dagger}_{\bm{k}}~c'^{\dagger}_{\bm{k}}~b'^{\dagger}_{\bm{k}}~a'_{-\bm{k}}~c'_{-\bm{k}}~
b'_{-\bm{k}}]. \label{eq:Yops}
\end{equation}

An analytic solution for the matrix $\mathsf{Q}$ that satisfies
all three conditions in (\ref{eq:maestro3}) is found to be
\begin{equation} \label{eq:Qfull}
\mathsf{Q}= \left [ \begin{array}{cc}
\mathsf{Q}_{11} & \mathsf{Q}_{12}\\
\mathsf{Q}_{12} & \mathsf{Q}_{11}\\
\end{array} \right], \nonumber
\end{equation}
where
\begin{equation} \label{eq:Q11mat}
\mathsf{Q}_{11}\!\!=\!\!\left [ \begin{array}{ccc}
0 & \frac{-\xi_1}{\sqrt{2A_1(1+\phi_1-A_1)}} & 0\\
\frac{-\xi_2}{\sqrt{2A_2(1+\phi_2-A_2)}} & 0 & 0\\
0 & 0 & \frac{-\theta}{\sqrt{2A_3(1+\theta-A_3)}}\\
\end{array} \right], \nonumber
\end{equation}
\begin{equation} \label{eq:Q12mat}
\mathsf{Q}_{12}= \left [ \begin{array}{ccc}
\sqrt{\frac{1+\phi_2}{2A_2}-\frac{1}{2}} & 0 & 0\\
0 & \sqrt{\frac{1+\phi_1}{2A_1}-\frac{1}{2}} & 0\\
0 & 0 & \sqrt{\frac{1+\theta}{2A_3}-\frac{1}{2}}\\
\end{array} \right], \nonumber
\end{equation}
and
\begin{displaymath}
\begin{array}{l l}
A_1 & = \sqrt{(1+\phi_1)^2-\xi_1^2}, \\
A_2 & = \sqrt{(1+\phi_2)^2-\xi_2^2}, \\
A_3 & = \sqrt{1+2\theta}.
\end{array}
\end{displaymath}

Knowing the transformation matrix $\mathsf{Q}$ we can then
determine the representation of the $\bm{L}$ and $\bm{S}$
operators in terms of the basis $Y$ of normal operators as
follows. Using (\ref{eq:szed}) the Fourier-transformed spin
operator components $S_\nu(\bm{k})$, with $\nu=z, +$ or $-$, can
be written in the generic form $S_\nu(\bm{k})=\mathcal{S}_\nu X$,
where $\mathcal{S}_{\nu}$ is a row vector of c-numbers, for
example $\mathcal{S}_z=[0~0~f_3~0~0~f_3]$. With respect to the
basis of normal operators, the Fourier-transformed spin operator
components become $S_\nu(\bm{k})=\mathcal{S}_\nu \mathsf{Q}Y$,
with similar expressions for the the orbital components
$L_{\nu}(\bm{k})$. Subsequently, we can evaluate all matrix
elements in (\ref{eq:int}) and thus calculate the neutron
scattering cross-section in Sec.~S6.

\section{S4. Wavefunctions in Applied Field and Mean Field Approximation}
\label{mean-field}

In this section we outline the derivation of the single-ion states
in the presence of an externally-applied magnetic field and
exchange interactions, treated within a mean-field approximation
following \cite{chen_prb}. Focusing on a single site, the relevant
Hamiltonian including the single-ion terms (\ref{eq:Hfirstap}) and
the exchange interactions (\ref{eq:ex}) is
\begin{equation} \label{eq:Hwex}
\mathcal{H}=\mathcal{H}_{\rm cf}+\mathcal{H}_{\rm SO}+\mu_B\bm{B}
\cdot (\bm{L} + 2\bm{S})+\sum_{\bm A}^{}J_{2}\bm{S} \cdot
\bm{S}_{\bm A},
\end{equation}
where the sum in the last term extends over all vectors ${\bm A}$
linking a \fep ion with its 12 NN on the same FCC sublattice. In
the mean-field approximation the spin operators $\bm{S}_{\bm A}$
are replaced by their expectation value $\langle \bm{S} \rangle$,
assumed to be the same for all sites, i.e. we search for
self-consistent solutions for the ground state of the single-site
mean-field Hamiltonian
\begin{equation} \label{eq:Hmf}
\mathcal{H}_{\rm mf}=\mathcal{H}_{\rm cf}+\mathcal{H}_{\rm
SO}+\mu_B\bm{B} \cdot (\bm{L} + 2\bm{S})+12J_{2}\bm{S} \cdot
\langle \bm{S} \rangle.
\end{equation}
We assume the ansatz
\begin{equation} \label{eq:ansatz}
\langle \bm{S} \rangle=m \bm{\hat{B}},
\end{equation}
where the field-induced spin polarization is along the applied
field direction, denoted by the unit vector $\bm{\hat{B}}$. Using
an assumed value for the spin polarization $m$ we diagonalize
(\ref{eq:Hmf}) to find the ground state wavefunction $\psi_0$ and
determine the expectation value of the spin polarization for that
state
\begin{equation} \label{eq:newm}
m'=\langle \psi_0 | \bm{S} \cdot \bm{\hat{B}} | \psi_0 \rangle
,\nonumber
\end{equation}
and search numerically for a self-consistent solution $m'=m$. For
this solution we the determine the wavefunctions $\psi_{0-3}$ of
the four lowest energy eigenstates of the mean-field Hamiltonian
(\ref{eq:Hmf}). If the field is applied along a cubic axis the
spin operator expansions in terms of bosons have the simplified
forms in (\ref{eq:szed}) and the dispersion relations can be
obtained analytically as listed in (\ref{eq:wdisp}). For a general
field direction the spin operator expansions (\ref{eq:szed})
generalize to contain up to three creation and annihilation terms
each, and the $3\times3$ matrices $A_{\bm k}$ and $B_{\bm k}$ in
(\ref{eq:Bmat}) have in general all elements non-zero. In this
case we numerically diagonalize the Hamiltonian matrix in
(\ref{eq:par}) to deduce the dispersion relations
$\hbar\omega_{1-3}(\bm{k})$ and the transformation matrix
${\mathsf{Q}}$ to obtain the neutron cross-section.

\begin{figure}[htbp]
\includegraphics[width=0.48\textwidth]{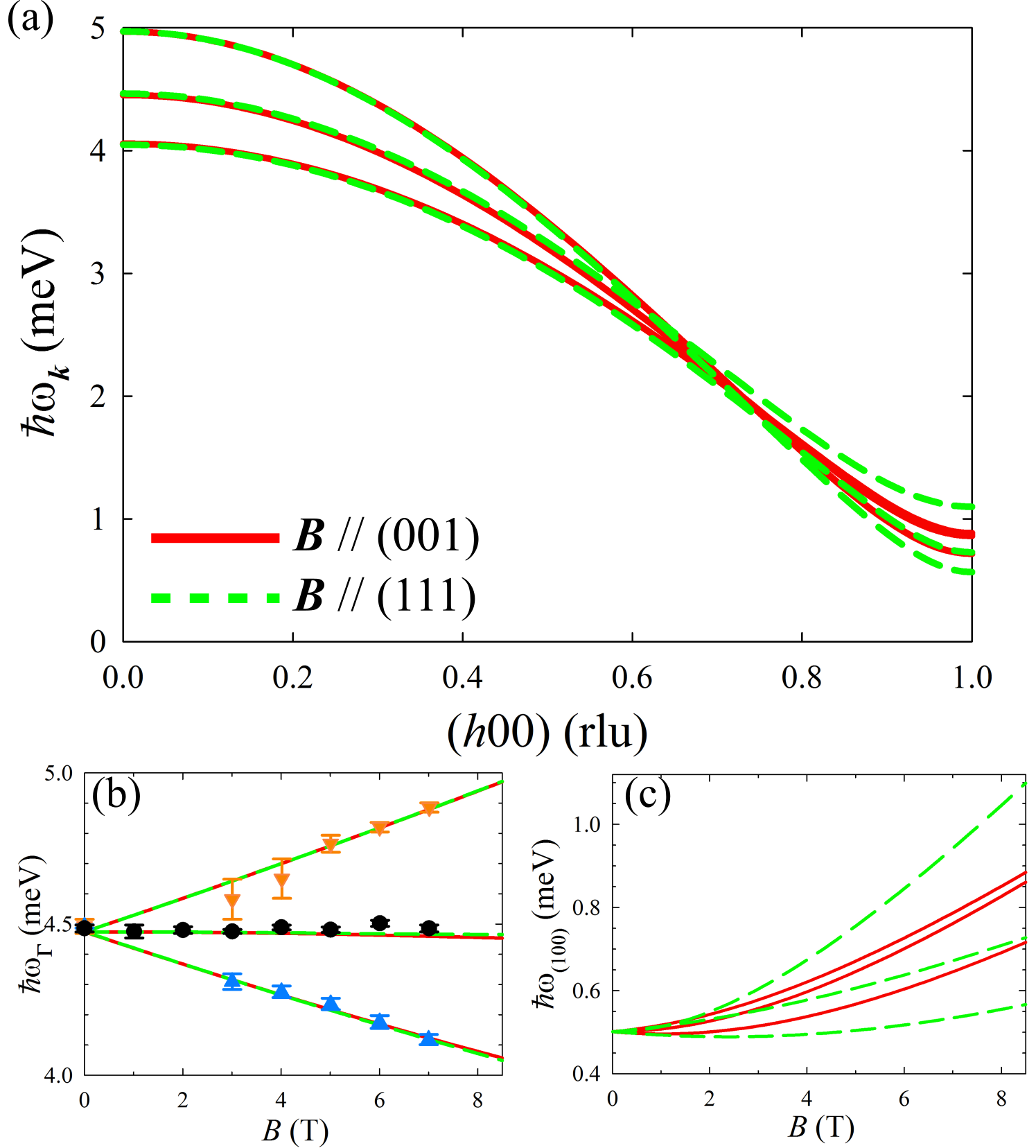}
\caption[]{(Color online) a) Dispersion relations along the
$(100)$ direction in reciprocal space for a magnetic field
$B=8.5$~T applied along (001) (solid red lines) and (111) (dashed
green lines), respectively. b) Splitting of excitation energies at
the zone center is linear in field and independent of direction,
in agreement with experimentally-measured THz transition energies
on powder samples (data points from \cite{laurita}). c) In
contrast. excitation energies at the softening point (100) show a
noticeable dependence on the applied field direction (solid red
lines for (001) and dashed green for (111).}\label{fig:fielddep}
\end{figure}

In the limit of small magnetic fields the behavior is independent
of the applied field direction \cite{ish}, however for the
magnitude fields used in the present study there is a significant
dependence of the triplon energies on the field orientation with
respect to the cubic axes. This anisotropy ultimately originates
in the fact that the crystal-field interaction $\mathcal{H}_{\rm
cf}$ in (\ref{eq:Hcfexap}) has only cubic, not spherical symmetry.
To illustrate this effect we plot in Fig.~\ref{fig:fielddep} the
dispersion relations along the high-symmetry (100) direction for a
magnetic field $B=8.5$~T applied along the cubic (001) axis (red
solid lines) and along the diagonal (111) direction (green dashed
lines), respectively. For both field directions the three-fold
degeneracy of the spin-orbital triplons is lifted resulting in
three non-degenerate modes. The field-dependence of the excitation
energies at the zone center ($\Gamma$-point) is plotted in
Fig.~\ref{fig:fielddep}b). Here the splitting is approximately
linear in field, independent of the field direction and moreover
the calculation is in quantitative agreement with no adjustable
parameters with the observed splitting of transition lines seen in
THz experiments on a powder sample (data points from
\cite{laurita}). At the mode softening wavevector (100) the
behavior is very different, non-linear in field, and the energies
depend strongly on the applied field direction as illustrated in
Fig.~\ref{fig:fielddep}c). It is the field behavior at those
wavevectors that is probed in the low-energy INS signal in
Figs.~3a-c) and 4a).

Finally we note that the application of a magnetic field leads to
a mixing between the zero-field states shown in Fig.~\ref{fig:cf},
inset) with the consequence that transitions become allowed
between the ground state $\psi_0$ and other higher energy states
derived from the $E$-doublet, in addition to transitions to the
first three excited states $\psi_{1,2,3}$. Specifically, this
mixing allows transitions between the ground state and high energy
states originating from the $\Gamma_3$ doublet in
Fig.~\ref{fig:cf} (inset). However, the INS data in applied
magnetic field shown in Fig.~3 is restricted to the region of low
to intermediate energy transfers when only transitions
$\psi_0\rightarrow\psi_{1,2,3}$ contribute, so the approximations
used in deriving the dispersion relations and intensities using
the three-flavor pseudo-boson method in Sec.~S3 are still expected
to be applicable.

%%%%%%%%%%%%%%%%%%%%%%%%%%%%%%%%%%%%%%%%%%%%%%%%%%%%%%%%%%%
\section{S5. Background Subtraction using detailed balance}
\label{Background}
%%%%%%%%%%%%%%%%%%%%%%%%%%%%%%%%%%%%%%%%%%%%%%%%%%%%%%%%%%%
In this section we outline the procedure used to estimate the
non-magnetic background contribution to subtract from the measured
INS data to obtain the pure magnetic signal. The method uses a
measured low-temperature data set, where magnetic signal is
expected to be present only on the positive energy side, and a
data set measured at relatively high temperatures, in the
paramagnetic phase, where a weaker, diffuse, magnetic signal is
expected to be present on both the positive and negative energy
sides. The relative intensities between the positive and negative
energy transfer for a given wavevector transfer $\bm{Q}$ are
related by the principle of detailed balance for the dynamical
structure factor \cite{squires}
\begin{equation}
S(-\bm{Q},-\hbar\omega) = e^{-\frac{\hbar \omega}{k_{\rm B} T}}
S(\bm{Q},\hbar\omega). \label{eq:detailedbalance}
\end{equation}
Formally, this is a consequence of the effect of time-reversal on
the dynamical structure factor, whereas physically it expresses
the fact that the intensity for a given process which transfers
energy $\hbar \omega$ to the neutron is exactly the same as for
the reverse process (when the neutron transfers energy $\hbar
\omega$ to the system) multiplied by a Boltzmann factor. It is
seen that in the limit of $T \rightarrow \infty$ both processes
are equally likely and contribute symmetrically to the intensity
profile. This principle applies regardless of the potential
responsible for the scattering. Eq.~(\ref{eq:detailedbalance})
implies the same Boltzmann factor relation between the
spherically-averaged structure factors $S(Q,\pm\hbar\omega)$, as
relevant for a powder INS experiment.

\begin{figure}[!htbp]
\includegraphics[width=0.48\textwidth]{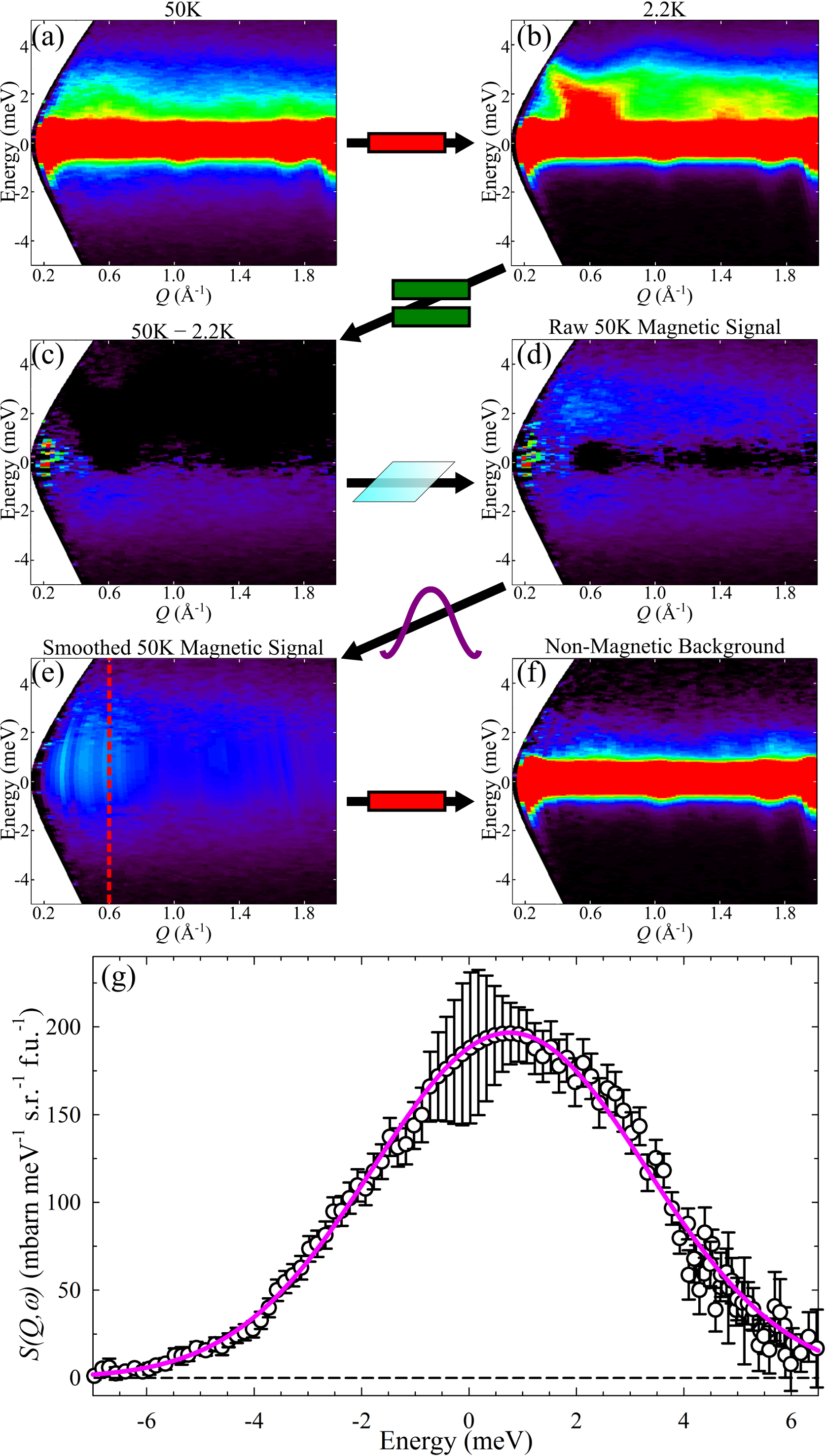}
\caption[]{(Color online) Background subtraction procedure: the
base temperature (2.2~K) raw data in (b) is subtracted from the
high temperature (50~K) raw data in (a) to obtain (c), which
contains pure paramagnetic scattering on the negative energy side.
The $\omega<0$ data in (c) is then `reflected' onto the $\omega>0$
side using (\ref{eq:detailedbalance}), then interpolated over the
elastic line region to obtain the estimated pure paramagnetic
scattering in (e), which is then subtracted from the raw data in
(a) to obtain the estimated non-magnetic background in (f). (g)
Energy scan at $Q=0.6(1)$\AA$^{-1}$ through the estimated pure
magnetic signal at 50~K in (e). Solid line is a fit to an
empirical functional form that satisfies the detailed balance
relation (\ref{eq:detailedbalance}) and is described in the text.
Dashed horizontal line emphasizes that plotted intensities are
background subtracted.} \label{fig:detailedbalance}
\end{figure}

The application of the principle of detailed balance to estimate
the non-magnetic background proceeds as follows; at very low
temperatures (2.2~K in the experiments outlined in the main text,
Fig.~\ref{fig:detailedbalance}(a)), the inelastic scattering is
concentrated almost entirely on the $\omega>0$ side of the
dynamical structure factor profile, as there are very few
thermally excited levels within the system able to transfer energy
{\it to} the neutron. As the temperature increases, the scattering
intensity spreads to the $\omega<0$ side as excited states become
thermally populated within the system. Assuming that the
contribution of magnetic scattering on the $\omega<0$ side (in
practice, below the elastic line) is negligible at base
temperature, by subtracting this intensity profile from a high
temperature data set (in practice, 50~K was found to be high
enough, Fig.~\ref{fig:detailedbalance}(b)) one can achieve an
estimate for the intensity of magnetic scattering processes that
transfer energy to the neutron at 50~K, this subtraction is shown
in Fig.~\ref{fig:detailedbalance}(c). However,
(\ref{eq:detailedbalance}) shows that the intensity on the
negative $\omega$ side is related to that on the positive $\omega$
side at the same $Q$ via the Boltzmann factor. Thereby, through
`reflecting' this negative $\omega$ intensity profile about the
elastic line taking account of the Boltzmann factor in
(\ref{eq:detailedbalance}), one arrives at an estimate of the high
temperature (50~K) magnetic scattering intensity, shown in
Fig.~\ref{fig:detailedbalance}(d). The `reflection' of the
magnetic signal works well for finite energy transfers away from
the elastic line, but is not applicable in the very close vicinity
of the elastic line where the signals to be subtracted between the
two data sets are very large and so extracting small differences
is not sufficiently reliable and/or there could be additional
scattering contributions with a distinct temperature dependence,
see the clear non-smooth behavior very close to the elastic line
in Fig.~\ref{fig:detailedbalance}(d). In this case we interpolate
the paramagnetic scattering intensity in the region covering the
close vicinity of the elastic line by assuming a smooth variation
of the diffuse scattering signal between the negative and positive
energy sides to obtain the plot in
Fig.~\ref{fig:detailedbalance}(e). This is illustrated in the
energy scan in Fig.~\ref{fig:detailedbalance}(g). The points below
$-1$~meV are from the subtraction 50~K minus 2.2~K data, points
above $1.5$~meV are obtained via `reflection', and points
in-between are interpolated. The solid line in the figure is a fit
to the functional form $I(\omega)=f(\hbar\omega/k_{\rm B}T)
G(\omega)$, where $f(x)=x/(1-e^{-x})$ and $G(\omega)$ is a
Gaussian of adjustable width centered at $\omega=0$. This
parametrization was chosen as i) it satisfies the detailed balance
principle in (\ref{eq:detailedbalance}),  ii) it converges at
$T\rightarrow\infty$ to a smooth profile centered at zero energy,
as expected for diffuse paramagnetic scattering, and iii)
empirically it appears to be a good parametrization of the
observed diffuse scattering, as shown by the comparison in
Fig.~\ref{fig:detailedbalance}(g). The estimated pure magnetic
signal at high temperature in Fig.~\ref{fig:detailedbalance}(e) is
then subtracted from the raw data in panel (a) to obtain the
estimated non-magnetic background plotted in panel (f), this in
turn is then subtracted from the low-temperature data in panel (b)
to obtain the pure magnetic signal plotted in Fig.~2(a).
%
%%%%%%%%%%%%%%%%%%%%%%%%%%%%%%%%%%%%%%%%%%%%%%%%%%%%%%%%%%
\section{S6. Powder-averaged Neutron Scattering intensity}
\label{cross-section}
%%%%%%%%%%%%%%%%%%%%%%%%%%%%%%%%%%%%%%%%%%%%%%%%%%%%%%%%%%
The inelastic neutron scattering intensity including polarization
and magnetic form factors is \cite{squires}
\begin{eqnarray}
&\mkern-144mu  S(\bm Q, \hbar\omega) =  (\gamma r_0)^2 f(Q)^2 \times \nonumber \\
&~~~~~~~~~~\underset{\alpha,\beta=x,y,z}{\sum}
\left(\delta_{\alpha \beta}-\frac{Q_\alpha Q_\beta}{Q^2} \right)
S^{\alpha \beta}(\bm Q, \hbar\omega), \label{eq:scat}
\end{eqnarray}
where $(\gamma r_0)^2=290.6$ mbarns/sr is a conversion factor
bringing the intensity into absolute units of
mbarns/meV/sr/formula unit, and $f(Q)$ is the magnetic form factor
for Fe$^{2+}$ ions. Here $Q_{x,y,z}$, are the components of the
wavevector transfer $\bm Q$ along the Cartesian $x,y,z$ axes.
$S^{\alpha \beta}(\bm Q, \hbar\omega)$ contain the dynamical
correlations for all possible transitions from an initial state,
$| i \rangle$ to a final state $| f \rangle$ given by
\begin{eqnarray}
S^{\alpha \beta}(\bm{Q},\hbar\omega) & = &\sum_{i,f} p_i \langle i
| L_{\alpha}+2S_{\alpha} | f \rangle \langle f |
L_{\beta}+2S_{\beta} | i \rangle \times \nonumber \\
& & \delta(\hbar \omega+ E_i -E_f),
\label{eq:magscatfunccrysfield}
\end{eqnarray}
where $p_i$ is the probability of the system initially being in
state $i$, $E_f-E_i$ is the energy transfer for the $i\rightarrow
f $ transitions, and the approximation $g_S = 2$ has been used. At
base temperature only the ground state is populated, $| i \rangle$
then corresponds to the product of $\psi_0(\bm{r})$ states at
every site $\bm{r}$ in the lattice, and the final states $| f
\rangle$ correspond to one-triplon states created by the normal
pseudo-boson operators $a'^{\dagger}_{\bm{k}}$,
$b'^{\dagger}_{\bm{k}}$ and $c'^{\dagger}_{\bm{k}}$ in
(\ref{eq:Yops}), with the dispersion relations
$\hbar\omega_{1,2,3}$ given in (\ref{eq:wdisp}).

We note that the dynamical correlations for the spin-orbital
singlet state have previously been calculated by treating the
exchange $J_2$ within a random-phase-approximation formalism
\cite{chen_prb}. Here we have provided an alternative approach by
deriving directly the dispersion relations in the presence of
exchange interactions via pseudo-boson triplon operators and
deriving explicitly the neutron scattering structure factor (via
the transformation to normal triplon operators) for both zero and
applied magnetic field.

For zero magnetic field the cross-section (\ref{eq:scat}) was
numerically averaged over a spherical distribution of orientations
of $\bm{Q}$ in order to obtain the orientational-averaged
intensity as a function of momentum $Q=|\bm{Q}|$ and energy
transfer, $S(Q, \hbar\omega)$, and this was directly compared with
the measured INS powder data in Fig.~2b). In a finite applied
magnetic field the dispersion relations (and neutron
cross-section) depend on the applied field direction with respect
to the cubic axes (as discussed in Sec.~S4), so in this case a
more elaborate averaging is required to reflect the fact that the
powder contains a spherically-uniform distribution of sample grain
orientations with respect to the instrument frame, but all grains
have the magnetic field applied along a fixed direction with
respect to the instrument frame. Since in the experimental
geometry used the (vertical) magnetic field was perpendicular to
the (horizontal) scattering plane of the detectors ($\bm{B} \perp
\bm{Q}$), the appropriate powder cross-section is obtained by
averaging the single-crystal cross-section (\ref{eq:scat}) over a
uniform distribution of wavevectors ${\bm Q}$ on a sphere of
radius $Q$ and choosing a uniform random direction of the magnetic
field in the plane normal to ${\bm Q}$. This method was used to
calculate the INS spectrum in Figs.~3d-f) and 4a)(dashed lines).

%%%%%%%%%%%%%%%%%%%%%%%%%%%%%%%%%%%%%%%%%%%%%%%%%%%%%%%%%%
\section{S7. Sample Preparation}
\label{sampleprep}
%%%%%%%%%%%%%%%%%%%%%%%%%%%%%%%%%%%%%%%%%%%%%%%%%%%%%%%%%%
%
\begin{figure}[htbp]
\includegraphics[width=0.46\textwidth]{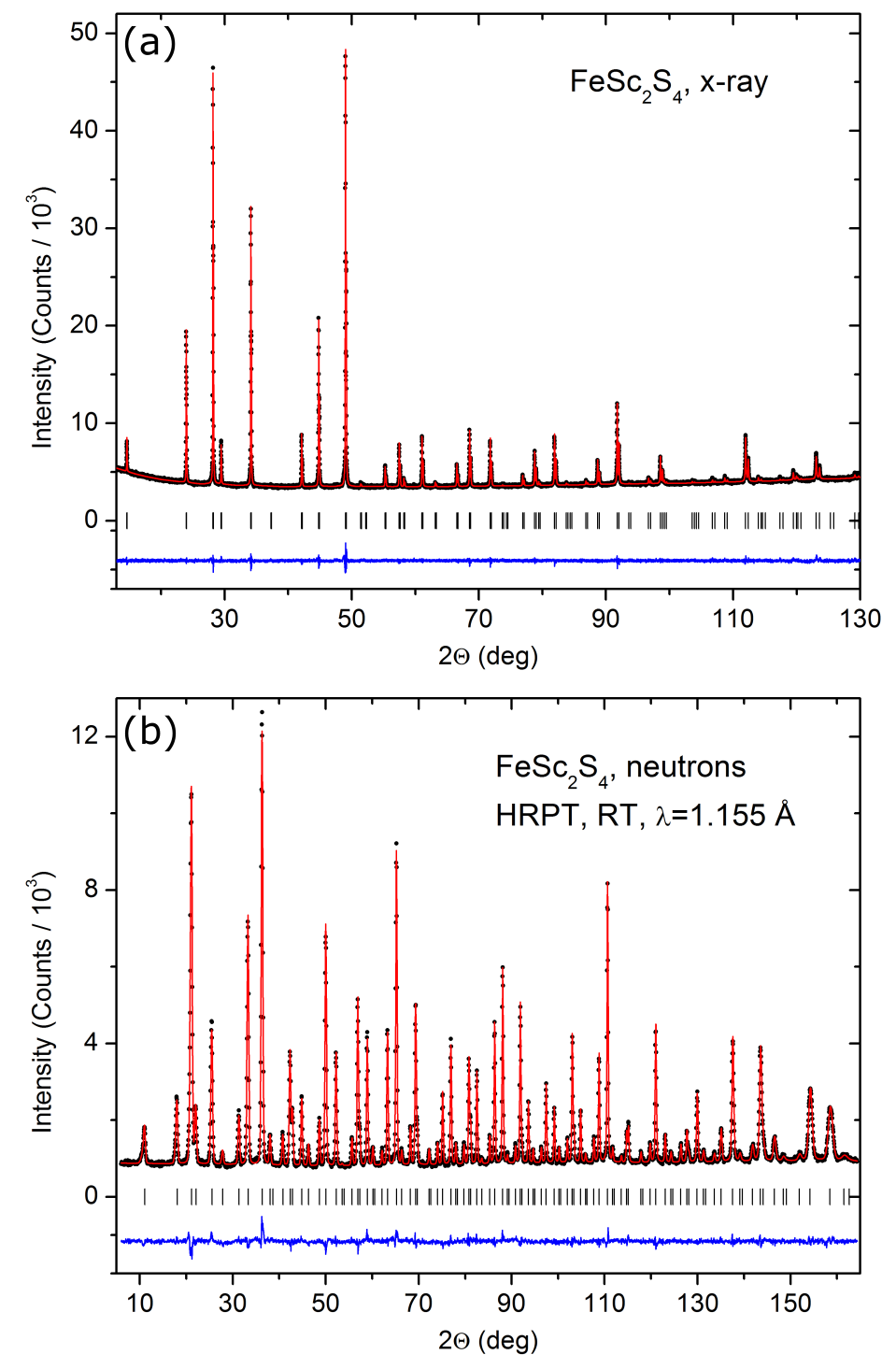}
\caption[]{(Color online) Rietveld refinement fit of \fesc\ crystal structure parameters at room temperature with data from (a) a Bruker D8 powder diffractometer and (b) the HRPT neutron powder diffractometer. Experimental points, calculated profile and the difference curve are shown. The ticks below the graph indicate the calculated positions of the diffraction peaks.}
\label{fig:refine}
\end{figure}
\begin{figure}[htbp]
\includegraphics[width=0.48\textwidth]{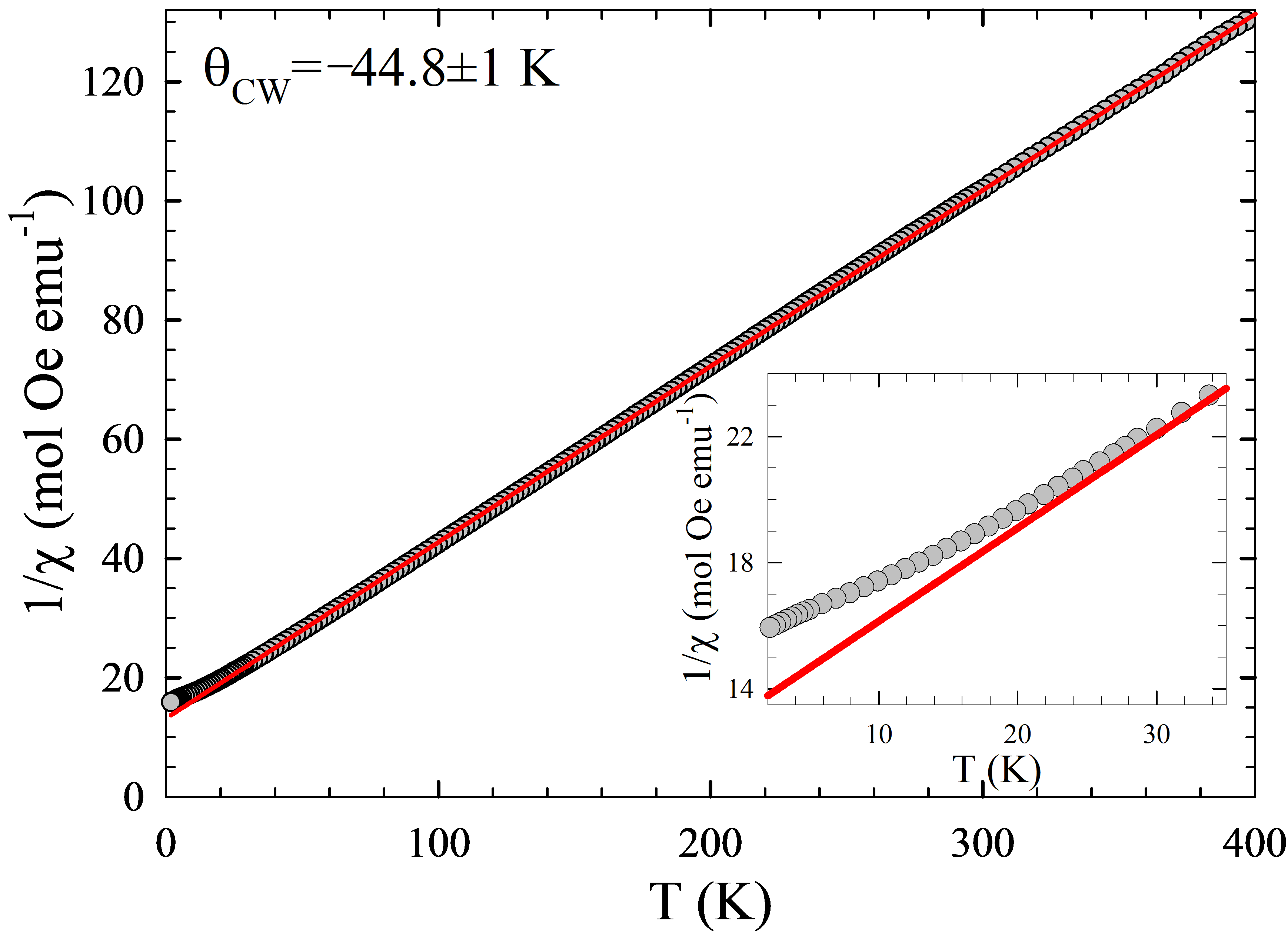}
\caption[]{(Color online) Inverse magnetic susceptibility of \fesc
powder as a function of temperature (gray circles) in an applied
field of $H=10^4$ Oe. The thick solid line is a fit to a Curie-Weiss
form $\chi=C/(T-\Theta_{\rm CW})$ giving $\Theta_{\rm CW}=-44.8\pm1$~K. Inset: zoom-in of the low temperature region
showing deviation from the linear behavior and the absence of any
sharp features indicative of a transition to magnetic order.}
\label{fig:squid}
\end{figure}
Polycrystalline \fesc\ was prepared by solid state synthesis from
the elements: Fe (99.99\%), Sc (99.9\%), and S (99.999\%).
Starting materials were loaded into quartz ampoules under Argon
atmosphere, then pumped to 10$^{-2}$~mbar and closed. After first
firing at 900$^\circ$C for 150~h the mixture was reground, pressed
into pellets, again closed within an ampoule and fired at the same
temperature. To reach full reaction, the sintering procedure was
repeated several times (up to 7 cycles). The samples after each
cycle were checked by SQUID magnetometry and XRD measurements. To
optimize the Fe:Sc:S ratio to the stoichiometric one, additional
heat treatments in vacuum and sulfur atmosphere at the last cycles
were performed. The composition of the sample was measured
by wave-length-dispersive X-ray electron-probe microanalysis (WDS
EPMA, Cameca SX50). The data were averaged over points measured on
15 different single-crystalline grains of about 40~$\mu$m in
diameter. The obtained composition was Fe 1.006(19) Sc 2.000(33) S
3.977(29) and corresponds to perfect stoichiometry (numbers in the
brackets give the standard deviations). 
% comments from Vladimir
%XRD studies of the final product by conventional
%powder diffraction (STADI-P, STOE \& Cie) did not find any
%non-reacted impurities or binary sulfides. The Rietveld analysis
%of the XRD pattern did not reveal essential inversion for cations
%between A and B sites.
%

The quality of our powder sample was controlled by x-ray and neutron powder diffraction. The x-ray powder data were collected with a Bruker D8 powder diffractometer (Cu K$_\alpha$$_{1,2}$ radiation) in an angular range between 4$^\circ$ and 130$^\circ$ in 2-theta. A profile-matching refinement shown in Fig \ref{fig:refine}(a) indicates a single phase material with no impurities. Neutron powder diffraction data were collected on the HRPT neutron powder diffractometer \cite{HRPT} at room temperature with $\lambda$=1.155 \AA\ neutrons in the angular range 4$^\circ$-165$^\circ$. It confirmed the phase purity of the material and allowed for a precise refinement of its crystal structure parameters. A Rietveld refinement carried out on this same neutron diffraction dataset Fig \ref{fig:refine}(b) also allowed for the refinement of possible cation disorder over the two cation sites in the structure, i.e. the distribution of the Fe and Sc cations in a compound with a nominal composition FeSc$_2$S$_4$ over the 8a(1/8,1/8,1/8) and 16d(1/2,1/2,1/2) sites, to be nominally occupied by solely iron and scandium, respectively. Even though the difference in the bound neutron scattering lengths for Fe and Sc (9.45 and 12.29 fm) is not very large, the relative simplicity of crystal structure in combination with a rather short wavelength -- thus covering a sufficiently broad Q-range, up to almost 11 \AA$^{-1}$ -- allows for rather precise refinement results. Given the perfect stochiometry of our sample, we assume iron and scandium are distributed in the ratio 1:2 over these two sites, and the level of disorder is parametrised by $0 \le x \le 1$ with Fe$_{1-x}$ Sc$_{x}$ occupying the 8a, and Sc$_{2-x}$ Fe$_{x}$ residing at the 16d sites. The case $x=0(1)$ represents a perfectly uninverted (inverted) structure. The resulting refinement yields a value of x=0.028(8) signifying an extremely low level of inversion. The refined coordinate of sulphur residing in the 32e($\varrho$,$\varrho$,$\varrho$) position is $\varrho$=0.25528(8) which is also quite a typical value for the AB$_2$O$_4$ compounds with spinel structures.

Results of magnetic susceptibility measurements (SQUID, MPMS-5, Quantum Design) are
shown in Fig.~\ref{fig:squid} and reveal a linear dependence of
the inverse susceptibility on temperature over the range $\approx$
20-400~K, in agreement with previous reports \cite{fritsch}. No
evidence for magnetic ordering was found down to the lowest
temperature probed, 1.8~K. We note that in contrast to the smooth susceptibility curve in Fig.~\ref{fig:squid}, studies of off-stoichiometry samples of Fe$_{1.06}$Sc$_{1.94}$S$_4$ show a clear anomaly at low temperatures as characteristic of the onset of long-range antiferromagnetic order \cite{tsurkanprep}. In contrast, for the powder sample studied here no such anomalies are present. Furthermore $\mu$SR data down to 1.5 K (not shown) indicated only a smooth relaxation without clear oscillations and neutron diffraction could not detect evidence for magnetic Bragg peaks, consistent with the absence of long-range magnetic order in the present samples.

As alluded to in the main text, recently Ref.~\cite{plumb} appeared reporting evidence for marginal
magnetic order in samples synthesized using a different protocol,
suggesting an extreme sensitivity to the synthesis route. We also noted that there are three main physical factors that could lead to such a discrepancy; {\it Vacancies}, {\it Site disorder}, and {\it Off-stoichiometry}. {\it Vacancies} at the A-site lead to randomly distributed absences in the diamond lattice of Fe$^{2+}$ ions, thus affecting the finely balanced frustration between NNN sites. Those at the B-site may also lead to a modulation of superexchange interactions as Sulphur ligands are displaced to compensate strains in the structure. However, experiments on samples deliberately synthesised with (up to 5\%) vacancies at the Fe sites have been shown to have similar magnetic and thermodynamic properties as those with the ideal crystal structure \cite{tsurkanprep}, suggesting that a small density of such absences is not detrimental. A-B {\it site disorder} is a common occurrence in spinels, and with the similar ionic radii of Sc$^{3+}$ and Fe$^{2+}$, great care must be taken in the synthesis of \fesc\ to minimise such disorder. {\it Off-stoichiometry} would also deeply affect the low temperature properties by introducing ionic species other than Fe$^{2+}$, Sc$^{3+}$ and S$^{2-}$ into the lattice. In particular, for the case of a surplus of Fe, one could presume, for example, the introduction of Fe$^{3+}$ ions into the lattice to preserve charge neutrality. Each of these carry an orbitally non-degenerate $S$=5/2 magnetic moment that could easily order when coupled by exchange interactions to the other magnetic ions in the lattice. To study these effects, we have synthesized samples with deliberate off-stoichiometry (e.g. the Fe$_{1.06}$Sc$_{1.94}$S$_4$ mentioned above) and find that those with a surplus of Fe do indeed show very different behaviour from pure \fesc . Concretely, magnetic susceptibility measurements on those Fe-rich samples show a deviation between field cooled and zero field cooled data as well as, crucially, the presence of magnetic order at low temperature \cite{tsurkanprep}.

\end{document}